\documentclass[floatfix,showpacs,showkeys,aps,prd,twocolumn]{revtex4-2}
\usepackage{lineno}
\usepackage{amsmath}
\usepackage{amssymb}

\usepackage{float}
\usepackage{tikz}
\usepackage{lipsum,adjustbox}
\usepackage{graphicx}
\usepackage[caption = false]{subfig}
\usepackage{dcolumn, multirow, booktabs, siunitx, rotating, array}
\usepackage{bm}
\usepackage{gensymb}

\usepackage{tabstackengine}
\usepackage[inline]{enumitem}

\setstackEOL{\cr}
\setstackgap{L}{\normalbaselineskip}

\let\oldequation\equation
\let\oldendequation\endequation

\renewenvironment{equation}
  {\linenomathNonumbers\oldequation}
  {\oldendequation\endlinenomath}

\newcolumntype{L}[1]{>{\raggedright\arraybackslash}p{#1}}
\newcolumntype{C}[1]{>{\centering\arraybackslash}p{#1}}
\newcolumntype{R}[1]{>{\raggedleft\arraybackslash}p{#1}}

\usepackage{hyperref}
\hypersetup{
 pdftitle={},
 pdfauthor={},
 pdfsubject={},
 pdfkeywords={},
 pdfstartview={},
 bookmarksopen=true, breaklinks=true, debug=true,
 colorlinks=true, linkcolor=blue, citecolor=red, urlcolor=blue,
 hyperfigures=true
}

\usepackage{todonotes}
\newcommand{\BESIIIorcid}[1]{\href{https://orcid.org/#1}{\hspace*{0.1em}\raisebox{-0.45ex}{\includegraphics[width=1em]{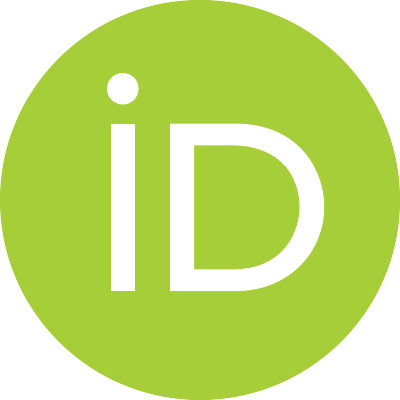}}}}

\begin{document}

\title{\boldmath Strong and electromagnetic amplitudes and direct $CP$ and isospin asymmetries in the decays $J/\psi \to K^0_SK^+\pi^-+\mathrm{c.c.}$}

\author{
M.~Ablikim$^{1}$\BESIIIorcid{0000-0002-3935-619X},
M.~N.~Achasov$^{4,c}$\BESIIIorcid{0000-0002-9400-8622},
P.~Adlarson$^{83}$\BESIIIorcid{0000-0001-6280-3851},
X.~C.~Ai$^{89}$\BESIIIorcid{0000-0003-3856-2415},
C.~S.~Akondi$^{31A,31B}$\BESIIIorcid{0000-0001-6303-5217},
R.~Aliberti$^{39}$\BESIIIorcid{0000-0003-3500-4012},
A.~Amoroso$^{82A,82C}$\BESIIIorcid{0000-0002-3095-8610},
Q.~An$^{78,65,\dagger}$,
Y.~H.~An$^{89}$\BESIIIorcid{0009-0008-3419-0849},
M.~S.~Anderson$^{39}$\BESIIIorcid{0009-0008-1550-2632}
% ...
\\
\vspace{0.2cm}
(BESIII Collaboration)\\
\vspace{0.2cm} {\it
$^{1}$ Institute of High Energy Physics, Beijing 100049, People's Republic of China\\
% ...
$^{4}$ Budker Institute of Nuclear Physics SB RAS (BINP), Novosibirsk 630090, Russia\\
% ...
$^{31}$ INFN Sezione di Ferrara, (A)INFN Sezione di Ferrara, I-44122, Ferrara, Italy; (B)University of Ferrara, I-44122, Ferrara, Italy\\
% ...
$^{39}$ Johannes Gutenberg University of Mainz, Johann-Joachim-Becher-Weg 45, D-55099 Mainz, Germany\\
% ...
$^{65}$ State Key Laboratory of Particle Detection and Electronics, Beijing 100049, Hefei 230026, People's Republic of China\\
% ...
$^{78}$ University of Science and Technology of China, Hefei 230026, People's Republic of China\\
% ...
$^{82}$ University of Turin and INFN, (A)University of Turin, I-10125, Turin, Italy; (B)University of Eastern Piedmont, I-15121, Alessandria, Italy; (C)INFN, I-10125, Turin, Italy\\
$^{83}$ Uppsala University, Box 516, SE-75120 Uppsala, Sweden\\
% ...
$^{89}$ Zhengzhou University, Zhengzhou 450001, People's Republic of China\\
\vspace{0.2cm}
$^{\dagger}$ Deceased\\
% ...
$^{c}$ Also at the Novosibirsk State University, Novosibirsk, 630090, Russia\\
% ...
}}

\date{\today}

\begin{abstract}
  Using $e^+e^-$ annihilation data collected at 26 center-of-mass energy points between 3000.00 and $3119.88~\text{MeV}$ with the BESIII detector, corresponding to a total integrated luminosity of about $440.7~\text{pb}^{-1}$, we study the production cross section line shape of the $e^+e^-\to K_S^0 K^+\pi^-+\mathrm{c.c.}$ process.
  A partial wave analysis is used to isolate the subprocesses $e^+e^-\to\bar K^0 K^*(892)^0+\mathrm{c.c.}$ and $e^+e^-\to K^+ K^*(892)^-+\mathrm{c.c.}$ with $K^*(892)^0\to K^+ \pi^-$ and $K^*(892)^-\to K_S^0 \pi^-$, respectively. The line shape fits determine the relative magnitude $\mathcal{C}$ and phase $\phi_{\gamma,3g}$ between the strong and electromagnetic amplitudes, together with the corresponding branching fractions $\mathcal{B}$. There is a twofold ambiguity in the relative phase determinations for two of the final states analyzed. Consequently, there are two equally possible sets of results on relative phase.
  For $(\phi_{\gamma,3g},\mathcal{C},\mathcal{B})$, the neutral mode has two solutions, $((147.0\pm 11.7)^\circ,\,3.80\pm 0.26,\,(4.62\pm 0.19)\times 10^{-3})$ and $((-146.0\pm 11.6)^\circ,\,3.85\pm 0.25,\,(4.80\pm 0.19)\times 10^{-3})$, whereas the charged mode has the single solution $((180.2\pm 35.2)^\circ,\,28.66\pm 2.68,\,(6.41 \pm 0.25)\times10^{-3})$. The neutral mode phase differs from $90^\circ$ by $4.8\sigma$, demonstrating that strong-electromagnetic orthogonality is not universal. The corresponding branching fractions agree with the world averages with substantially improved precision. A simultaneous fit to the neutral and charged channels gives that the neutral-to-charged strong contribution ratio $\mathcal{R}^{3g}_{K^*\bar K}=1.062\pm 0.070$ or $1.090\pm 0.063$, respectively, both consistent with isospin symmetry. The direct $CP$ asymmetry analysis shows no evidence for $CP$ violation. The uncertainties include both statistical and systematic contributions.
\end{abstract}

\maketitle

\section{INTRODUCTION}
Quantum chromodynamics (QCD) is well tested in the asymptotically free short-distance region and in the confining long-distance region~\cite{QCD1,QCD2,QCD3,QCD4}. However, the dynamics in the transition region where a heavy $c\bar c$ pair hadronizes into light-hadron final states remains poorly understood~\cite{QCDheavyquark,Quarkonia}. At the $J/\psi$ mass region, the narrow 93 keV $J/\psi$ decay width supports the use of perturbative QCD, and the leading strong-decay mechanism is Okubo-Zweig-Iizuka (OZI)-suppressed annihilation into three gluons. Perturbative QCD (pQCD) holds the point that the amplitude of $e^+e^-$ annihilations into hadrons at very high center-of-mass (c.m.) energies ($\sqrt s$) is expected to be real~\cite{nuFF,heliSele}. However, an unexpected $90^\circ$ phase between the strong three-gluon and electromagnetic (EM) one-photon amplitudes from $J/\psi$ has been observed~\cite{ppphase,NNphase}. SU(3) flavor symmetry-based phenomenological models~\cite{SU3-1-inter,SU3-2-jpsietac,SU3-3-mo} can reproduce such a phase, treating the $J/\psi$ as an SU(3) singlet and introducing an additional isospin-violating strong contribution, while the underlying dynamical origin of a large absorptive contribution at the $J/\psi$ scale is still under debate.

Several mechanisms have been discussed in the literature. An additional imaginary amplitude can be introduced by the mixing between the $J/\psi$ and a vector glueball~\cite{gluonball1,gluonball2,gluonball3}, and the hadronic loop effect of the final-state interactions (FSI)~\cite{FSI1,FSI2}. In particular, an approximately channel-independent imaginary amplitude can be accommodated if the $J/\psi$ mixes strongly with a vector glueball close in mass through an off-diagonal self-energy term. In this scenario, the vector glueball is expected to have a much larger hadronic width than the $J/\psi$, while its leptonic width is strongly suppressed. Such mixing has been proposed as a possible explanation of the long-standing ``$\rho\pi$ puzzle,'' the violation of the pQCD 12\% rule, defined as $\frac{\mathcal{B}(\psi(2S)\to h)}{\mathcal{B}(J/\psi\to h)} = \frac{\mathcal{B}(\psi(2S)\to e^+e^-)}{\mathcal{B}(J/\psi\to e^+e^-)} = 12.7\%$, which was first observed by the Mark-II Collaboration in the $\rho\pi$ and $K^*\bar{K}$ modes~\cite{rhopiMarkII}. Alternatively, it has been argued that one may account for the relevant phase structure using OZI-violation mechanisms alone~\cite{OZIamp}. In a gluonic-annihilation picture, once the running of the strong coupling is treated properly, the dispersive representation of the OZI-violating amplitude can reproduce the correct sign of $\phi$--$\omega$ mass mixing and implies a predominantly imaginary amplitude in the $J/\psi$ mass region. This provides a natural explanation of the empirical observation that the OZI-violating amplitude around the $\phi$ resonance is fully imaginary, without invoking explicit nonsinglet intermediate states or additional glueball-mixing assumptions.

In another picture, the $J/\psi$ couples to off-shell $D^{(*)}\bar D^{(*)}$ intermediate states~\cite{FSIopechar1,FSIopechar2}, which subsequently transition into light hadrons through strong interactions. Since the initial coupling to charmed hadrons does not suffer from OZI suppression, the resulting dispersive loop amplitudes can be comparable in magnitude to the three-gluon contribution and naturally carry nontrivial strong phases. Their interference with the OZI-violating amplitude can, therefore, play a decisive role in shaping both the decay rates and their channel dependence. Moreover, the cancellation of charged and neutral charmed meson loops in the exact isospin limit is spoiled by realistic mass splittings of $m_{D^{(*)+}}\neq m_{D^{(*)0}}$, providing a natural dynamical mechanism for enhanced isospin-violating transitions of $\psi(2S)\to\pi^0 J/\psi$ relative to $\psi(2S)\to\eta J/\psi$~\cite{pi0jpsi}. Therefore, isospin symmetry can also be broken in $J/\psi$ decays. These strong phases are also relevant to $CP$ studies. Within the Standard Model, purely strong and QED charmonium decays are expected to exhibit negligible direct $CP$ violation; therefore, precise comparisons between charge-conjugate channels provide clean null tests. More generally, hadronic phase differences are required for direct $CP$ violation once an additional amplitude carrying a weak phase is present, and, thus, can test new physics effects and give constraints in other heavy hadron systems.

The decay channel of $J/\psi\to K_S^0 K^+\pi^-$ is selected to study the decay of $J/\psi\to K^*(892)\bar K$. This vector-pseudoscalar final state is one of the classic channels discussed in connection with the long-standing ``$\rho\pi$ puzzle'' and has also attracted interest as a testing ground for helicity selection expectations in exclusive charmonium decays. Throughout this paper, charged conjugations are always implied apart from discussing the $CP$ test. With modern BESIII statistics and an energy scan across the $J/\psi$ region, we update the relevant branching fractions (BFs) and perform a partial wave analysis (PWA) to isolate the $K^*(892)\bar K$ contributions, in parallel with a cross section line shape analysis to access the coherent interplay of strong and EM amplitudes. In addition, the corresponding continuum production $e^+e^-\to K^*(892)\bar K$ exhibits a striking isospin pattern, experimental measurements in both charmonium and bottomonium energy regions report large ratios, up to $\sim 20$~\cite{Belle:2013hkg,CLEO:2004vxx}. This is in tension with simple SU(3) symmetry models' estimation that incorporate charge factors and kinematics, which typically predict only a modest enhancement, up to $\sim 4$~\cite{Carlsmith:1987qa,Lu:2018obb}.

In this paper, we study the cross section line shape by analyzing $e^+e^-$ collision data samples taken at 26 c.m. energies ($\sqrt s$) between 3000.00 and $3119.88~\text{MeV}$, which correspond to a total integrated luminosity of $440.7~\text{pb}^{-1}$, and perform a PWA across all energies. For data samples at each energy point, the c.m. energies are measured with the beam energy measurement system, which is based on measuring the energies of Compton backscattered photons~\cite{Abakumova:2011rp}; the integrated luminosities are measured by analyzing the large-angle Bhabha scattering and diphoton events~\cite{wangyadi,jlumi,jpsinumber}.

\begin{figure*}[htbp]\centering
  \includegraphics[width=0.325\textwidth]{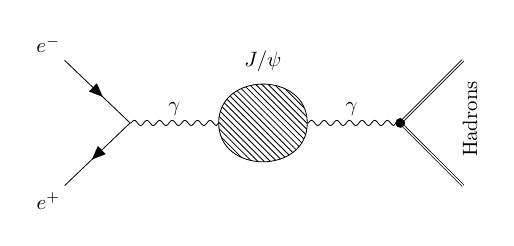}
  \includegraphics[width=0.325\textwidth]{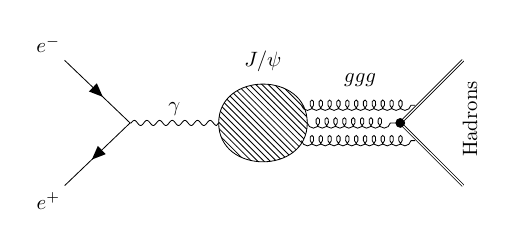}
  \includegraphics[width=0.325\textwidth]{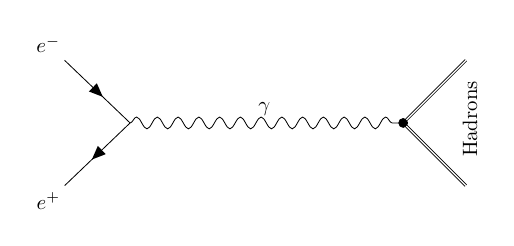}
  \caption{The Feynman diagrams for $e^+e^-\to$ hadrons: (left) $J/\psi$ EM decay via one virtual photon, (middle) $J/\psi$ strong decay via gluons, and (right) continuum production via a virtual photon.}
  \label{fig:fey}
\end{figure*}

Near the narrow resonance $J/\psi$, exclusive cross sections mainly consist of the coherent sum of three amplitudes~\cite{jpsidecay}: the strong amplitude mediated by three gluons $\mathcal{A}_{3g}$, the EM amplitude mediated by one photon $\mathcal{A}_\gamma$, and continuum production via a virtual photon $\mathcal{A}_\text{cont}$, as shown in Fig.~\ref{fig:fey}. The total Born cross section for $e^+e^-\to f$ is expressed as a coherent sum of resonance and continuum amplitudes:
\begin{equation}
  \sigma_\text{Born}^{f}(s) \propto \left|\mathcal{A}^{f}_\text{cont}(s)+\mathcal{A}^{f}_\gamma(s)+\mathcal{A}^{f}_{3g}(s) \right|^2.
\end{equation}

Assuming that the three contributions share the same effective final-state dynamics, the Born cross section for the $e^+e^-\to K_S^0 K^+\pi^-$ process and its subprocesses around $J/\psi$ is written as~\cite{wangpingformular,wangyadiformular,Zlineshape,kedr}
\begin{widetext}
\begin{equation}\label{eq:xsformular}
  \sigma_\text{Born}^{f}=\frac{4 \pi \alpha^2}{3 s}\left|1+(1+\mathcal{C} e^{i \phi_{\gamma, 3g}}) \frac{s}{M} \frac{3 \sqrt{\Gamma^0_{e e} \Gamma^0_{\mu\mu}} / \alpha}{s-M^2+i M \Gamma}e^{i\phi_{\gamma,\text{cont}}}\right|^2 \mathcal{P}(s) \frac{\mathcal{F}^2}{s^n},
\end{equation}
\end{widetext}
where 
$\alpha$ is the fine-structure constant, $\Gamma_{ee}^0$ and $\Gamma_{\mu\mu}^0$ are the bare electronic and muonic widths, respectively, $M$ and $\Gamma$ are the mass and width of the $J/\psi$ resonance, respectively, $\mathcal{C}$ is the ratio $\frac{\left|\mathcal{A}_{3g}\right|}{\left|\mathcal{A}_\gamma\right|}$, and $\phi_{\gamma,3g(\rm cont)}$ is the relative phase between strong (continuum) and EM amplitudes. The $\phi_{\gamma, \text{cont}}$ was previously determined to be around zero in the analyses of the cross sections of $J/\psi\to \mu^+\mu^-$ and $J/\psi\to \eta\pi^+\pi^-$ at BESIII~\cite{wangyadi}. The factor $\mathcal{F}^2 / s^n$ parametrizes the squared effective timelike EM form factor for the exclusive final state $f$, with $n$ governing the energy dependence. Depending on the ``counting rule"~\cite{heliSele}, $n$ is set to be 3 for the two-body final states $\bar K^0 K^*(892)^0$ and $K^+ K^*(892)^-$. For the three-body final state $K_S^0 K^+ \pi^-$, we set $n = 5$~\cite{BaBar:2007ceh,Belle:2013hkg}. The symbol $\mathcal{P}(s)$ is the dimensionless phase-space (PHSP) factor. For the two-body subprocesses, it is written as
\begin{equation}
  \mathcal{P}=
  \left[
    \left(1-\frac{m_1^2}{s}-\frac{m_2^2}{s}\right)^2
    -\frac{4m_1^2m_2^2}{s^2}
  \right]^{3/2},
\end{equation}
where $(m_1,m_2)=(m_{\bar K^0},m_{K^*(892)^0})$ or $(m_{K^+},m_{K^*(892)^-})$. For the three-body final state $K_S^0K^+\pi^-$, the PHSP factor is evaluated as
\begin{equation}
  \begin{aligned}
  \mathcal{P}=&
  \frac{1}{128\pi^3s^2}
  \int_{(m_{K_S^0}+m_{K^+})^2}^{(\sqrt{s}-m_{\pi^-})^2}
  \\
  &\times
  \frac{
    \sqrt{\lambda(s_{12},m_{K_S^0}^2,m_{K^+}^2)
    \lambda(s,s_{12},m_{\pi^-}^2)}
  }{s_{12}}
  \,ds_{12},
  \end{aligned}
\end{equation}
where $s_{12}$ is the invariant mass squared of $K_S^0$ and $K^+$, and $\lambda(x,y,z)=x^2+y^2+z^2-2xy-2xz-2yz$.

Based on a $\chi^2$ line shape fit to the observed cross section, which accounts for the initial state radiation (ISR) and beam energy spread effects as detailed in Sec.~\ref{sec:xsfit}, we extract the Born cross sections and the relative magnitude and phase between the strong and EM amplitudes in $J/\psi \to K^0_SK^+\pi^-$. In combination with the PWA results, these parameters are extracted in its dominant subprocesses $J/\psi\to K^+ K^*(892)^-$ (charged channel) and $J/\psi\to\bar K^0 K^*(892)^0$ (neutral channel). Isospin breaking is quantified by the neutral-to-charged ratios at the Born level, cross sections, and BFs:
\begin{equation}
  \begin{aligned}
    &\mathcal{R}_\sigma = \frac{\sigma_{\rm Born}(e^+e^-\to\bar K^0 K^*(892)^0)}{\sigma_{\rm Born}(e^+e^-\to K^+ K^*(892)^-)},\\
    &\mathcal{R}_{K^*\bar{K}} = \frac{\mathcal{B}(J/\psi\to \bar K^0 K^*(892)^0)}{\mathcal{B}(J/\psi\to K^+ K^*(892)^-)}.\\
  \end{aligned}
\end{equation}
To separate the contributions from the strong and EM interactions, the total $J/\psi$ decay amplitude is written as $\mathcal{A}_\gamma + \mathcal{A}_{3g}= \mathcal{A}_\gamma (1 + \mathcal{C}e^{i\phi_{\gamma,3g}})$. The corresponding fractions of the strong and EM contributions to the total decay rate are then expressed as
\begin{equation}
  \begin{aligned}
  &f_{3g} = \frac{|\mathcal{A}_{3g}|^2}{|\mathcal{A}_{3g} + \mathcal{A}_\gamma|^2}
  = \frac{\mathcal{C}^2}{1 + 2\mathcal{C}\cos\phi_{\gamma,3g} + \mathcal{C}^2},\\
  &f_{\gamma} = \frac{|\mathcal{A}_\gamma|^2}{|\mathcal{A}_{3g} + \mathcal{A}_\gamma|^2}
  = \frac{1}{1 + 2\mathcal{C}\cos\phi_{\gamma,3g} + \mathcal{C}^2},
  \end{aligned}
\end{equation}
while the interference term is given by
\begin{equation}
  f_\text{int} = \frac{2\mathcal{C}\cos\phi_{\gamma,3g}}{1 + 2\mathcal{C}\cos\phi_{\gamma,3g} + \mathcal{C}^2}.
\end{equation}
For each channel $(i)$, we obtain $f_{3g}^{(i)}$ and $f_{\gamma}^{(i)}$ from the best-fit parameters. The neutral to charged ratios for the pure strong and pure EM parts are then
\begin{equation}
  \begin{aligned}
    &\mathcal{R}^{3g}_{K^*\bar{K}} = \mathcal{R}_{K^*\bar{K}}\times \frac{f_{3g}^{(K^*(892)^0)}}{f_{3g}^{(K^*(892)^-)}},\\
    &\mathcal{R}^{\gamma}_{K^*\bar{K}} = \mathcal{R}_{K^*\bar{K}}\times \frac{f_{\gamma}^{(K^*(892)^0)}}{f_{\gamma}^{(K^*(892)^-)}}.\\
  \end{aligned}
\end{equation}

Direct $CP$ null tests are performed using the Born cross sections:
\begin{equation}
A_{\sigma_{\rm Born}} =\frac{\sigma_{\rm Born}^{+}-\sigma_{\rm Born}^{-}}
{\sigma_{\rm Born}^{+}+\sigma_{\rm Born}^{-}},
\end{equation}
where the superscripts ``$+$" and ``$-$" denote the samples split by the unambiguous flavor tag provided by the charged kaon. For each channel, we determine asymmetries of the fit parameters:
\begin{equation}
  \begin{aligned}
  &A_{\mathcal{C}}=\frac{\mathcal{C}^{+}-\mathcal{C}^{-}}{\mathcal{C}^{+}+\mathcal{C}^{-}},\\
  &\Delta\phi_{\gamma,3g}=\phi_{\gamma,3g}^{+}-\phi_{\gamma,3g}^{-},\\
  &A_{\mathcal{B}}=\frac{\mathcal{B}^{+}-\mathcal{B}^{-}}{\mathcal{B}^{+}+\mathcal{B}^{-}}.\\
  \end{aligned}
\end{equation}
These observables probe possible $CP$ asymmetries in the relative magnitude, the phase, and the BF, respectively.

\section{BESIII EXPERIMENT AND MONTE CARLO SIMULATION}

The BESIII detector~\cite{BES3} records symmetric $e^+e^-$ collisions provided by the BEPCII storage ring~\cite{Yu:2016cof} in the c.m. energy range from 1.84 to 4.95~GeV, with a peak luminosity~($\mathcal{L}$) of $1.1 \times 10^{33}\;\text{cm}^{-2}\text{s}^{-1}$ achieved at $\sqrt{s} = 3.773\;\text{GeV}$. BESIII has collected large data samples in this energy region~\cite{bes3-white-paper,EcmsMea,EventFilter}. The cylindrical core of the BESIII detector covers 93\% of the full solid angle and consists of a helium-based multilayer drift chamber~(MDC), a plastic scintillator time-of-flight system~(TOF), and a CsI(Tl) EM calorimeter~(EMC), which are all enclosed in a superconducting solenoidal magnet providing a 1.0~T magnetic field. The solenoid is supported by an octagonal flux-return yoke with resistive plate counter muon identification modules interleaved with steel. 
The charged-particle momentum resolution at $1~{\rm GeV}/c$ is $0.5\%$, and the ${\rm d}E/{\rm d}x$ resolution is $6\%$ for electrons from Bhabha scattering. The EMC measures photon energies with a resolution of $2.5\%$ ($5\%$) at $1$~GeV in the barrel (end-cap) region. The time resolution in the TOF barrel region is 68~ps, while that in the end-cap region was 110~ps. The end-cap TOF system was upgraded in 2015 using multigap resistive plate chamber technology, providing a time resolution of 60~ps, which benefits about 80\% of the data used in this analysis~\cite{etof}.

Simulated samples are produced with a {\sc geant4}-based~\cite{geant4} Monte Carlo (MC) package, which includes the geometric description of the BESIII detector and the detector response. The simulations model the beam energy spread and ISR in the $e^+e^-$ annihilations with the generator {\sc kkmc}~\cite{kkmc}. To estimate backgrounds, an inclusive MC sample is generated including both the production of the $J/\psi$ resonance and the continuum processes incorporated in {\sc kkmc}~\cite{kkmc}. All particle decays are modeled with {\sc evtgen}~\cite{evtgen} using BFs either taken from the Particle Data Group~(PDG)~\cite{pdg}, when available, or otherwise estimated with {\sc lundcharm}~\cite{lundcharm}. Final-state radiation from charged final-state particles is incorporated using the {\sc photos} package~\cite{photos}. The signal decays $J/\psi\to K_S^0 K^+\pi^-$, $J/\psi\to\bar K^0 K^*(892)^0$, and $J/\psi\to K^+ K^*(892)^-$ are generated with ConExc~\cite{conexc} using the helicity amplitude with parameters fixed to the results obtained by the PWA in this analysis.

\section{EVENT SELECTION}

Charged tracks detected in the MDC (except for those used for $K_S^0$ reconstruction) are required to be within a polar angle $(\theta)$ range of $\vert\!\cos\theta\vert<0.93$, where $\theta$ is defined with respect to the $z$ axis, which is the symmetry axis of the MDC. The distance of closest approach to the interaction point (IP) along the $z$ axis, $|V_z|$, must be less than 10 cm and in the transverse plane, $|V_{xy}|$, less than 1 cm. Particle identification (PID) for charged tracks combines measurements of the specific ionization energy loss in the MDC ($\text{d}E/\text{d}x$) and the flight time in the TOF to form likelihoods $\mathcal{L}(h)(h=K,\pi)$ for each hadron $h$ hypothesis. Charged kaons and pions are identified by comparing the likelihoods for the kaon and pion hypotheses, $\mathcal{L}(K)>\mathcal{L}(\pi)$ and $\mathcal{L}(\pi)>\mathcal{L}(K)$, respectively.

Each $K_S^0$ candidate is reconstructed from two oppositely charged tracks satisfying $|V_z|<20~\text{cm}$. The two charged tracks are assigned as $\pi^+\pi^-$ without imposing PID criteria. They are constrained to originate from a common vertex, requiring an invariant mass within $|M_{\pi^+\pi^-}-m_{K_S^0}| < 15~\text{MeV}/c^2$, where $m_{K_S^0}$ is the $K^0_{S}$ nominal mass~\cite{pdg}. The decay length of the $K_S^0$ candidate is required to be greater than twice the vertex resolution away from the IP. The quality of both primary and secondary vertex fits is ensured by a requirement on the $\chi^2~(\chi^2<100)$.

To suppress background, a four-constraint (4C) kinematic fit imposing energy-momentum conservation is carried out under the $e^+e^-\to K_S^0 K^+ \pi^-$ hypothesis. If multiple $K_S^0$ candidates are present in an event, the one with the longest $K_S^0$ decay length is retained, and candidate events are required to satisfy $\chi_\text{4C}^2<200$. To remove gamma conversion Bhabha background, we require the opening angle between the two pions from $K_S^0$ decay to be greater than 30$^\circ$, denoted as $\theta_{\pi^+\pi^-}>30^\circ$.

Potential background sources are studied using inclusive $J/\psi$ MC sample (INC-MC) and exclusive $e^+e^-\to K_S^0K^+\pi^-\pi^0$, $e^+e^-\to K^+K^-\pi^+\pi^-$, and $e^+e^-\to 2(\pi^+\pi^-)$ MC samples after applying the same event selection criteria. The study of INC-MC shows a negligible background contribution (0.2\%), from $J/\psi\to\gamma\eta_c\to \gamma K_S^0 K^+ \pi^-$. The exclusive background MC samples are generated by ConExc using the respective cross sections~\cite{BaBarkkpipi, BaBarkskpipi0, BaBar4pi} with a PHSP model. The misidentification rate is less than 0.01\%, that is negligible, when normalized to the integrated luminosity.

Non-$K_S^0$ backgrounds are further examined using events in the $K_S^0$ sideband, defined as $ M_{\pi^+\pi^-} \in (0.40,\,0.46) \cup (0.54,\,0.60)~\text{GeV}/c^2$. After scaling mass window range, the event ratios between sideband and signal regions are 0.4\% in data and 0.2\% in signal MC sample, indicating a negligible background contribution. Consequently, the signal yield for $e^+e^-\to K_S^0 K^+ \pi^-$ is obtained simply by counting events. 

To improve the resolution of kinematic variables, a five-constraint kinematic fit is performed, in which, in addition to imposing energy and momentum conservation, $M_{\pi^+\pi^-}$ is constrained to the known $K_S^0$ mass~\cite{pdg}.

\section{PARTIAL WAVE ANALYSIS}\label{sec:pwa}

A PWA fit is performed on the surviving candidate events to identify the intermediate processes. The decay amplitude is constructed using the helicity amplitude formalism~\cite{Chungformalism, Richmanformalism}, and the full fit procedure is implemented based on an open-source framework called TF-PWA~\cite{tfpwajiangyi}.

\subsection{Helicity amplitude} 

\begin{figure*}[htbp]\centering
  \includegraphics[width=0.32\textwidth]{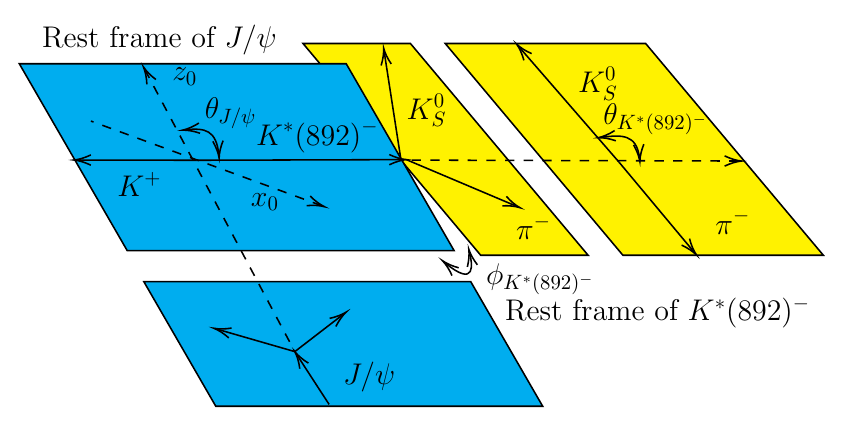}
  \includegraphics[width=0.32\textwidth]{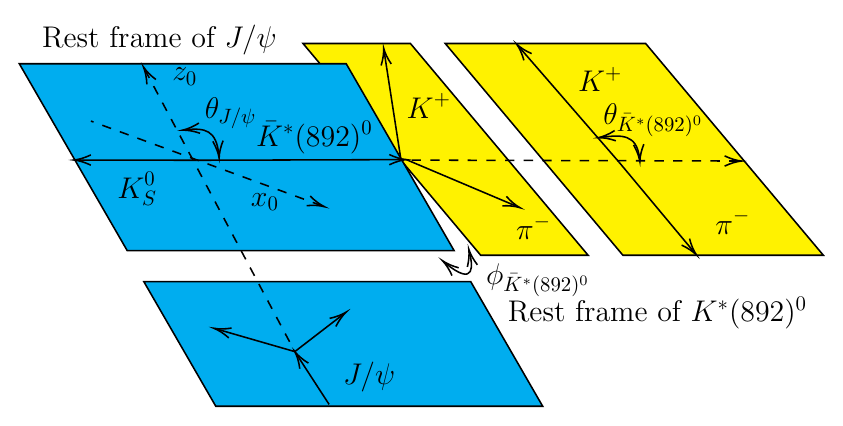}
  \includegraphics[width=0.32\textwidth]{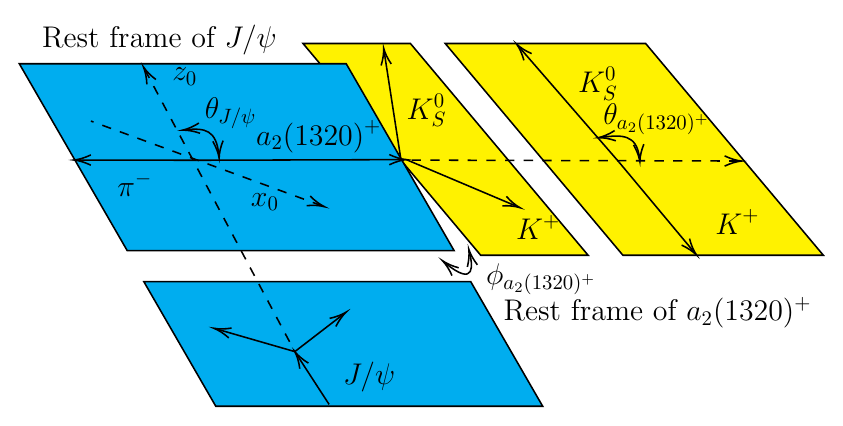}
  \caption{The corresponding helicity angle definitions for $K^*(892)^-$, $K^*(892)^0$, and $a_2(1320)^+$ are shown as examples.}
  \label{fig:heliangle}
\end{figure*}

To construct the full amplitude of $e^+e^-\to\gamma^{*}\to f$, the helicity formalism is used based on the isobar model describing the three-body decay as a two-step sequential quasi-two-body process. For each two-body process $0\to 1+2$, the helicity amplitude is written as
\begin{equation}
  \mathcal{M}_{\lambda_0, \lambda_1, \lambda_2}^{0 \to 1+2}=\mathcal{H}_{\lambda_1, \lambda_2}^{0 \to 1+2} D_{\lambda_0, \lambda_1-\lambda_2}^{J_0 *}(\phi, \theta, 0),
\end{equation}
where the helicity coupling $\mathcal{H}_{\lambda_1, \lambda_2}^{0 \to 1+2}$ is given by the $LS$ coupling formula \cite{Chung1993, Chung1997} along with barrier factor terms
\begin{widetext}
  \begin{equation}
    \mathcal{H}_{\lambda_1, \lambda_2}^{0 \to 1+2}=\sum_{l s} g_{l s} \sqrt{\frac{2 l+1}{2 J_0+1}}\left\langle l 0, s \delta \mid J_0, \delta\right\rangle\left\langle J_1 \lambda_1, J_2-\lambda_2 \mid s, \delta\right\rangle(\frac{q}{q_0})^l B_l^{\prime}(q, q_0, d),
  \end{equation}
\end{widetext}
where $g_{l s}$ is the partial wave amplitude, $J_{0,1,2}$ are the spins of the particles 0, 1, and 2, $\lambda_{1,2}$ are the helicities of the particles 1 and 2, and $\delta = \lambda_1 - \lambda_2$ is the helicity difference. Here, $q$ is the three-momentum modulus of particle 1 in the rest frame of particle 0, which is calculated as
\begin{equation}
  q=\frac{\sqrt{[m^2-(m_1+m_2)^2][m^2-(m_1-m_2)^2]}}{2 m},
\end{equation}
where $m$, $m_1$ and $m_2$ are the masses of the particles 0, 1, and 2, respectively. The normalization factor $q_0$ is calculated at the nominal resonance mass. The factor $B_l^{\prime}(q, q_0, d)$ is the reduced Blatt-Weisskopf factor \cite{rBW, centrifugal}, which is given by
\begin{widetext}
\begin{equation}
  \begin{aligned}
    & B_0^{\prime}(q, q_0, d)=1, \\
    & B_1^{\prime}(q, q_0, d)=\sqrt{\frac{1+(q_0 d)^2}{1+(q d)^2}},\\
    & B_2^{\prime}(q, q_0, d)=\sqrt{\frac{9+3(q_0 d)^2+(q_0 d)^4}{9+3(q d)^2+(q d)^4}},\\
    & B_3^{\prime}(q, q_0, d)=\sqrt{\frac{225+45(q_0 d)^2+6(q_0 d)^4+(q_0 d)^6}{225+45(q d)^2+6(q d)^4+(q d)^6}},\\
    & B_4^{\prime}(q, q_0, d)=\sqrt{\frac{11025+1575(q_0 d)^2+135(q_0 d)^4+10(q_0 d)^6+(q_0 d)^8}{11025+1575(q d)^2+135(q d)^4+10(q d)^6+(q d)^8}}.
  \end{aligned}
  \label{eq:Blatt-Weisskopf}
\end{equation}
\end{widetext}
In the Wigner $D$ function, $D_{\lambda_0, \lambda_1-\lambda_2}^{J_0*}(\phi, \theta, 0)$, $\phi$ and $\theta$ are helicity angles, as shown in Fig.~\ref{fig:heliangle}. The definitions can be found in Ref.~\cite{Wang2020}. In Eq.~\eqref{eq:Blatt-Weisskopf}, the radius of the centrifugal barrier $d$ is chosen as $d=0.73~\mathrm{fm}$, consistent with Ref.~\cite{BESpsipkketa}.

The amplitude of a complete chain is constructed as the product of each two-body amplitude and the resonant propagator $R$. In the sequential process $e^+e^-\to\gamma^*(\to J/\psi), \gamma^*\to R_{K_S^0 \pi^-}K^+, R_{K_S^0 \pi^-}\to K_S^0\pi^-$, the amplitude is written as
\begin{widetext}
\begin{equation}
    \mathcal{M}_{\lambda_{K_S^0}, \lambda_{K^+}, \lambda_{\pi^-}}^{\lambda_{\gamma^*}, R_{K_S^0 \pi^-}}=\sum_{\lambda_{R_{K_S^0 \pi^-}}}\mathcal{M}_{\lambda_{\gamma^*},\lambda_{R_{K_S^0 \pi^-}}, \lambda_{K^+}}^{\gamma^*\to R_{K_S^0 \pi^-}K^+} R_{R_{K_S^0 \pi^-}}(M_{K_S^0\pi^-}) \mathcal{M}_{\lambda_{R_{K_S^0 \pi^-}}, \lambda_{K_S^0}, \lambda_{\pi^-}}^{R_{K_S^0 \pi^-}\to K_S^0\pi^-}.
\end{equation}
\end{widetext}
For the decay proceeding through the $R_{K^+ \pi^-}$ intermediate state, the amplitude is expressed as
\begin{widetext}
  \begin{equation}
    \mathcal{M}_{\lambda_{K_S^0}, \lambda_{K^+}, \lambda_{\pi^-}}^{\lambda_{\gamma^*}, R_{K^+ \pi^-}}=\sum_{\lambda_{R_{K^+ \pi^-}}}\mathcal{M}_{\lambda_{\gamma^*},\lambda_{R_{K^+ \pi^-}}, \lambda_{K_S^0}}^{\gamma^*\to R_{K^+ \pi^-}K_S^0} R_{R_{K^+ \pi^-}}(M_{K^+\pi^-}) \mathcal{M}_{\lambda_{R_{K^+ \pi^-}}, \lambda_{K^+}, \lambda_{\pi^-}}^{R_{K^+ \pi^-}\to K^+\pi^-}.
  \end{equation}
\end{widetext}
For the decay via the $R_{K_S^0 K^+}$ intermediate states, the amplitude reads
\begin{widetext}
  \begin{equation}
    \mathcal{M}_{\lambda_{K_S^0}, \lambda_{K^+}, \lambda_{\pi^-}}^{\lambda_{\gamma^*}, R_{K_S^0 K^+}}=\sum_{\lambda_{R_{K_S^0 K^+}}}\mathcal{M}_{\lambda_{\gamma^*},\lambda_{R_{K_S^0 K^+}}, \lambda_{\pi^-}}^{\gamma^*\to R_{K_S^0 K^+}\pi^-} R_{R_{K_S^0 K^+}}(M_{K_S^0 K^+}) \mathcal{M}_{\lambda_{R_{K_S^0 K^+}}, \lambda_{K_S^0}, \lambda_{K^+}}^{R_{K_S^0 K^+}\to K_S^0 K^+}.
  \end{equation}
\end{widetext}
For the nonresonant (NR) decay, the amplitude is replaced by setting $R_{NR}(M_{K_S^0K^+})$ as unity, expressed as
\begin{equation}
  \mathcal{M}_{\lambda_{K_S^0}, \lambda_{K^+}, \lambda_{\pi^-}}^{\lambda_{\gamma^*}, NR}=\sum_{\lambda_{NR}}\mathcal{M}_{\lambda_{\gamma^*},\lambda_{NR}, \lambda_{\pi^-}}^{\gamma^*\to NR~\pi^-} \mathcal{M}_{\lambda_{NR}, \lambda_{K_S^0}, \lambda_{K^+}}^{NR\to K_S^0K^+}.
\end{equation}
The propagators of all resonances are described by the relativistic Breit-Wigner formula, written as
\begin{equation}
  R(m)=\frac{1}{m_0^2-m^2-i m_0 \Gamma(m)},
\end{equation}
with the mass-dependent width of
\begin{equation}
  \Gamma(m)=\Gamma_0(\frac{q}{q_0})^{2 l+1} \frac{m_0}{m} B_l^{\prime 2}(q, q_0, d).
\end{equation}

The full amplitude is the coherent sum of all possible resonances, given as
\begin{widetext}
\begin{equation}\label{eq:amptot}
  |\mathcal{M}|^2=\sum_{\Lambda,\Lambda^\prime}\rho_{\Lambda,\Lambda^\prime}\sum_{\lambda_{K_S^0}=0}\sum_{\lambda_{K^+}=0}\sum_{\lambda_{\pi^-}=0}\left|\mathcal{M}^{R_{K^+ \pi^-}}_{\lambda_{K_S^0}, \lambda_{K^+}, \lambda_{\pi^-}}+\mathcal{M}^{R_{K_S^0 \pi^-}}_{\lambda_{K_S^0}, \lambda_{K^+}, \lambda_{\pi^-}}+\mathcal{M}^{R_{K_S^0 K^+}}_{\lambda_{K_S^0}, \lambda_{K^+}, \lambda_{\pi^-}}\right|^2,
\end{equation}
\end{widetext}
where $\rho_{\Lambda,\Lambda^\prime}$ is the spin density matrix of the initial virtual photon.

To simplify the PWA fit, an overall resonance amplitude for each component term in Eq.~\eqref{eq:amptot}, such as those listed in Table~\ref{tab:pwasignificance}, can be derived to represent the overall magnitude and phase of the specific component. In each cascade process, one of the partial wave amplitudes $g_{ls}$ can be taken as reference amplitude with a fixed amplitude of constant 1. Hence, the relative magnitudes and phases of the other partial wave amplitudes in each cascade process are left free in the fit.

\subsection{Likelihood function construction and fit fraction} 

The probability density function (PDF) for a given event is constructed using the full amplitude as
\begin{equation}
  \text{PDF}(p_j)=\frac{\epsilon(p_j)\left|\mathcal{M}(p_j)\right|^2 R_3(p_j)}{\int \epsilon(p_j)\left|\mathcal{M}(p_j)\right|^2 R_3(p_j) \mathrm{d} p_j},
  \label{eq:PDF}
\end{equation}
where $\epsilon(p_j)$ is the detection efficiency parametrized in terms of the final four-momenta $p_j$ and $j$ refers to the different particles in the final states. $R_3(p_j)$ is the standard element of the three-body PHSP. The integration part in the denominator is calculated by the sum of the PHSP MC, e.g.,
\begin{equation}
  \int \epsilon(p_j)\left|\mathcal{M}(p_j)\right|^2 R_3(p_j) \mathrm{d} p_j \propto \frac{1}{N_{\mathrm{PHSP}}} \sum_{i \in \mathrm{PHSP}}\left|\mathcal{M}(p_i)\right|^2.
\end{equation}
The negative log likelihood (NLL) is formed as
\begin{equation}
  -\ln L=-\left[\sum_{i \in \mathrm{data}} \ln \text{PDF}(x_i)\right].
  \label{eq:NLL}
\end{equation}
After minimizing the NLL, the parameters' error matrix is calculated by the inverse Hessian matrix:
\begin{equation}
  V_{i j}^{-1}=-\frac{\partial^2 \ln L}{\partial \theta_i \partial \theta_j},
  \label{eq:hessian_matrix}
\end{equation}
where $\theta_i$ is the $i$th free parameter in the likelihood formula. This is the standard form of the error matrix in the maximum likelihood fit.

The fit fraction (FF) of each resonance state is calculated using
\begin{equation}
  \mathrm{FF}_i=\frac{\int\left|\mathcal{M}_i\right|^2 \mathrm{~d} \Phi}{\int\left|\sum_k \mathcal{M}_k\right|^2 \mathrm{~d} \Phi} \propto \frac{\sum_{j \in \operatorname{PHSP}}\left|\mathcal{M}_i(x_j)\right|^2}{\sum_{j \in \operatorname{PHSP}}\left|\sum_k \mathcal{M}_k(x_j)\right|^2},
\end{equation}
where $\mathcal{M}_i$ is the amplitude of the $i$th component and the integration is calculated by summing the PHSP MC events at the generator level. The statistical uncertainties for FFs are obtained with the standard error propagation. For the determination of the subprocess cross sections, we also calculate reconstruction level FFs, $\mathrm{FF}_i^{\rm rec}$, by replacing the generator level PHSP MC events with the reconstructed level PHSP MC events.

\subsection{Nominal fit results} 

The fit procedure includes all possible intermediate states from the PDG~\cite{pdg} that satisfy $J^{PC}$ conservation in the subsequent two-body decay. The PWA fit procedure starts by including $K^*(892)^0$, $K^*(892)^-$, $K_2^*(1430)^0$, $K_2^*(1430)^-$, and $a_2(1320)^+$ as the initial baseline solutions. The dominant $K^*(892)^0$ component is chosen as the reference channel, with its magnitude and phase fixed to one and zero, respectively. Hence, the relative complex couplings of the other partial wave amplitudes in each cascade process are left free. For the resonances, masses and widths are fixed to their PDG values~\cite{pdg}.

The statistical significance of each contribution is evaluated based on the change of the NLL value when including or excluding individual components, while taking into account the change in the number of degrees of freedom. The tested components include $K^{*}(1410)$, $K^{*}(1680)$, $K_3^{*}(1780)$, $K_2^{*}(1980)$, and $K_4^{*}(2045)$ for $K^*$ candidates; $\rho(1450)^+$, $\rho_3(1690)^+$, $\rho(1700)^+$, $\rho(1900)^+$, $\rho_3(1990)^+$, $\rho(2150)^+$, and $\rho_3(2250)^+$ for excited $\rho$ candidates; as well as the nonresonant $\mathcal{P}$-wave $NR_{1^-}$ and $\mathcal{D}$-wave $NR_{2^+}$ components in the $M_{K_S^0 K^+}$ spectrum. The above strategy is implemented individually on the experimental datasets collected at $\sqrt{s}=3080.00$ and $3096.20~\text{MeV}$, which are chosen for their high integrated luminosities and for representing distinct $e^+e^-$ annihilation regimes: the continuum-dominated process $\mathcal{A}_\text{cont}$ at $3080.00~\text{MeV}$ and the resonance region where $\mathcal{A}_\text{cont}$, $\mathcal{A}_\gamma$, and $\mathcal{A}_{3g}$ interfere at $3096.20~\text{MeV}$. The results show that the resonances $K_4^*(2045)^-$, $K^*(1680)^0$, and $K_3^*(1780)^0$ for the dataset collected at $\sqrt{s}=3080.00~\text{MeV}$ and the resonances $\rho(1450)^+$, $K^*(1680)^-$, $K^*(1680)^0$, $K_2^*(1980)^-$, and $K_2^*(1980)^0$, along with $NR_{1^-}$ for the dataset collected at $\sqrt{s}=3096.20~\text{MeV}$, have statistical significance larger than $5\sigma$, while none of the other tested contributions exceeds this threshold.

\begin{figure*}[htbp]\centering
	\includegraphics[width=0.99\textwidth]{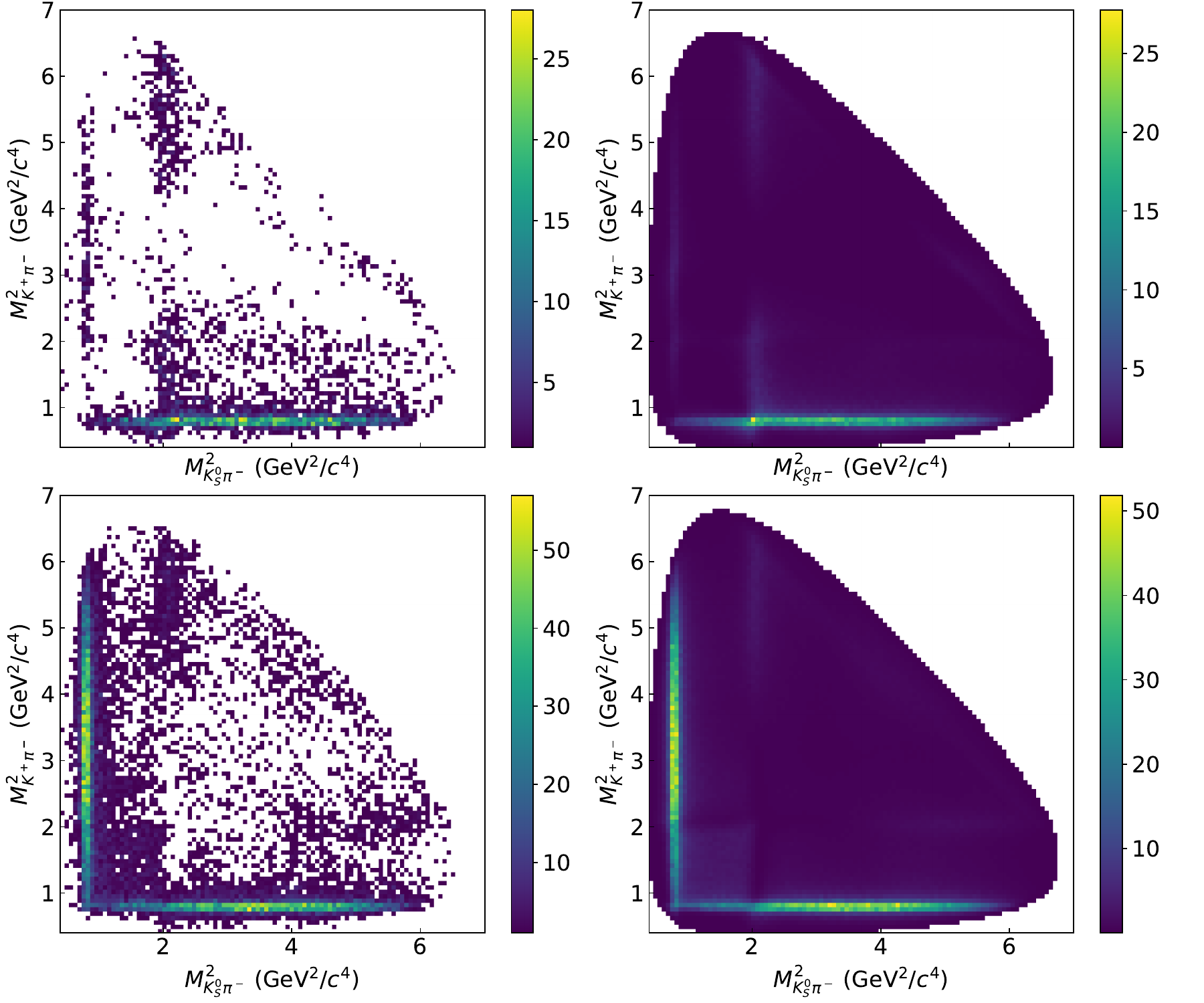}
	\caption{The Dalitz plots of $m^2_{K_S^0\pi^-}$ versus $m^2_{K^+\pi^-}$ in datasets (left) and fitted results (right) at $\sqrt s=$ 3080.00 MeV energy point (top) and $\sqrt s=$ 3096.20 MeV energy point (bottom).}
	\label{fig:dalitz}
\end{figure*}

\begin{figure*}[htbp]\centering
  \includegraphics[width=0.995\textwidth]{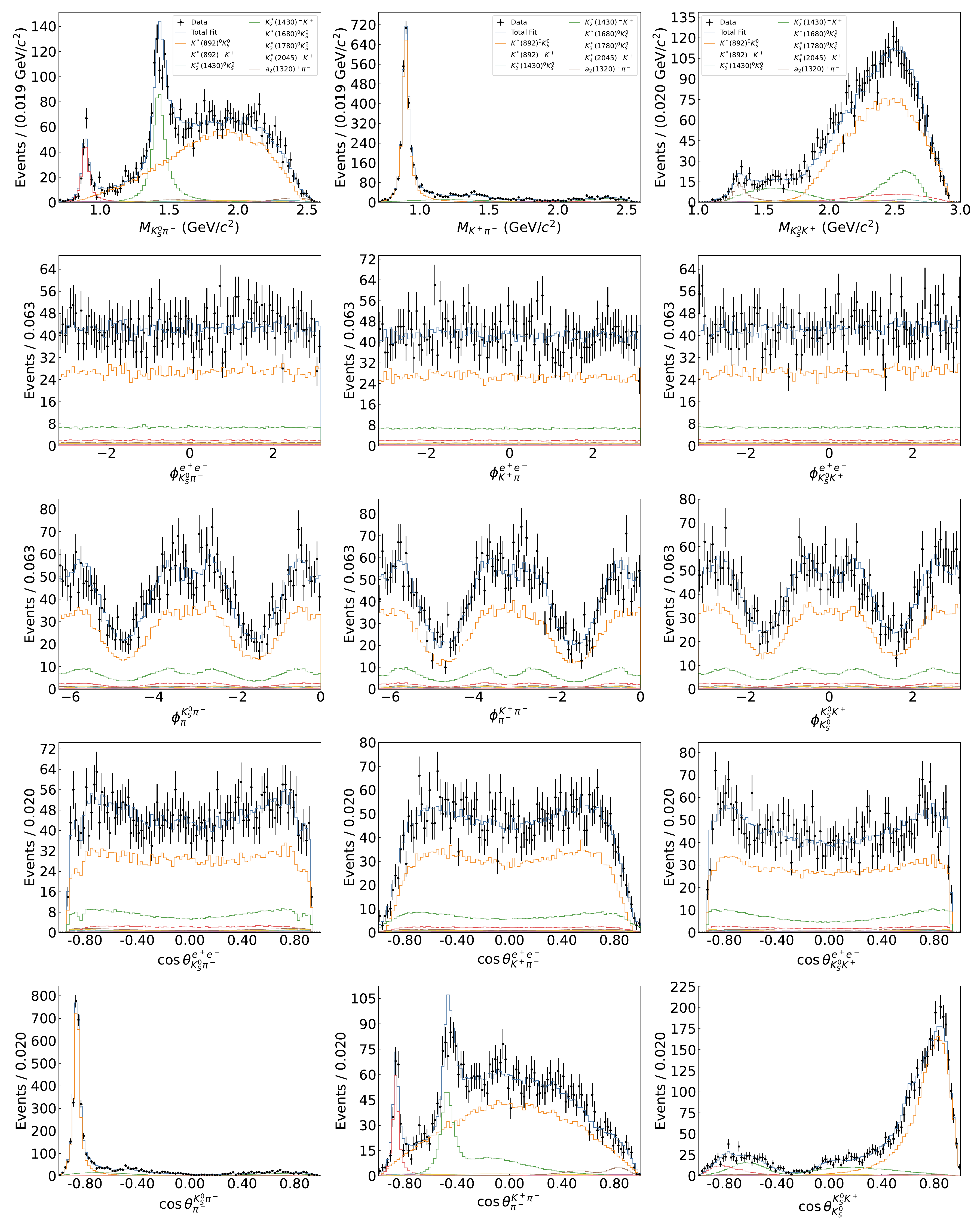}
  \caption{Projections of the PWA fit on mass spectra and helicity angles for the dataset taken at $\sqrt s=3080.00$ MeV.}
	\label{fig:pwaresult3080.0}
\end{figure*}

\begin{figure*}[htbp]\centering
  \includegraphics[width=0.995\textwidth]{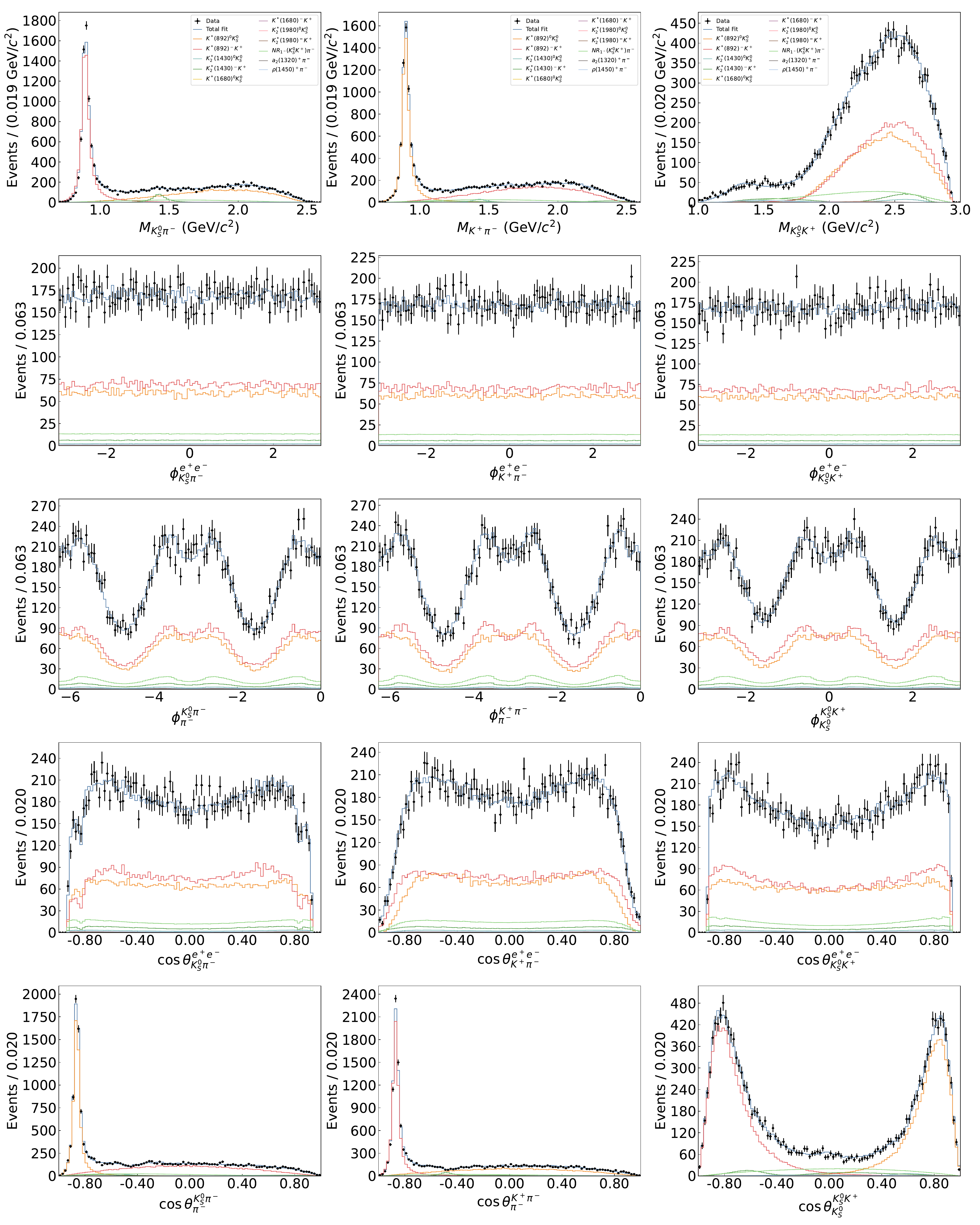}
  \caption{Projections of the PWA fit on mass spectra and helicity angles for the dataset taken at $\sqrt s=3096.20$ MeV.}
	\label{fig:pwaresult3096.2}
\end{figure*}

Therefore, the nominal components are determined to be the $K^*(892)^0$, $K^*(892)^-$, $K_2^*(1430)^0$, $K_2^*(1430)^-$, $K_4^*(2045)^-$, $K^*(1680)^0$, $K_3^*(1780)^0$, and $a_2(1320)^+$, for the dataset collected at $\sqrt{s}=3080.00~\text{MeV}$, and $K^*(892)^0$, $K^*(892)^-$, $K_2^*(1430)^0$, $K_2^*(1430)^-$, $\rho(1450)^+$, $K^*(1680)^-$, $K^*(1680)^0$, $K_2^*(1980)^-$, $K_2^*(1980)^0$, and $a_2(1320)^+$ as well as the nonresonant $NR_{1^-}$ component for the dataset collected at $\sqrt{s}=3096.20$ MeV, whose statistical significances are listed in Table~\ref{tab:pwasignificance}. The fractions of the interference parts can be found in Tables~\ref{tab:ff3080.0} and \ref{tab:ff3096.2}.
The Dalitz plots of data and fit results are shown in Fig.~\ref{fig:dalitz}, and other one-dimensional projections are shown in Figs.~\ref{fig:pwaresult3080.0} and \ref{fig:pwaresult3096.2}. For the resonances $K^*(892)^0$ and $K^*(892)^-$, further studies on the decay asymmetries and the relevant systematic uncertainties are performed.
The intermediate state configuration obtained from the scan at $\sqrt{s}=3080.00~\text{MeV}$ is adopted for datasets collected at $\sqrt{s}=3000.00$ to $3088.85~\text{MeV}$, while the configuration determined at $\sqrt{s}=3096.20~\text{MeV}$ is used for the remaining energy points. For each energy point, an independent PWA fit is performed with the corresponding fixed intermediate state set, and the complex couplings of all amplitudes are refitted.

\begin{table}[htbp]\centering
  \caption{The statistical significance values for each component in the nominal fit model.}
  \scalebox{1.0}{
    \begin{tabular}{ccccc}
      \toprule
      \multirow{2}{*}{Resonance} & \multicolumn{2}{c}{Significance ($\sigma$)}  \\
      \cmidrule(lr){2-3}
                            & $\sqrt s =3080.00$ MeV & $\sqrt s =3096.20$ MeV \\
      \midrule
      $K^*(892)^0$          & $>30$    & $>30$  \\
      $K_2^*(1430)^0$       & 12.0     & 19.7   \\
      $K^*(1680)^0$         & 5.3      & 6.4    \\
      $K_2^*(1980)^0$       & ...      & 6.6    \\
      $K_3^*(1780)^0$       & 7.5      & ...    \\
      $K^*(892)^-$          & 23.8     & $>30$  \\
      $K_2^*(1430)^-$       & $>30$    & 28.1   \\
      $K^*(1680)^-$         & ...      & 7.4    \\
      $K_2^*(1980)^-$       & ...      & 8.8    \\
      $K_4^*(2045)^-$       & 5.8      & ...    \\
      $a_2(1320)^+$         & 10.8     & 5.1    \\
      $\rho(1450)^+$        & ...      & 11.7   \\
      $NR_{K_S^0 K^+}(1^-)$ & ...      & 19.7   \\
      \bottomrule
    \end{tabular}}
  \label{tab:pwasignificance}
\end{table}

\begin{table*}[htbp]\centering
  \caption{The nominal FFs (in units of \%) of each component and the interference for the dataset taken at $\sqrt s=3080.00$ MeV. Total refers to the sum of the FFs without considering interference. The uncertainties are statistical only.}
  \resizebox{\linewidth}{!}{
    \begin{tabular}{l|cccccccc}
      \toprule
      Mode & $K^*(1680)^0$ & $K^*(892)^0$ & $K_2^*(1430)^0$ & $K_2^*(1430)^-$ & $K_3^*(1780)^0$ & $K_4^*(2045)^-$ & $K^*(892)^-$ & $a_2(1320)^+$ \\
      \midrule
      $K^*(1680)^0$ & $(2.21 \pm 0.50)\%$ & - & - & - & - & - & - & - \\
      $K^*(892)^0$ & $(-0.64 \pm 0.48)\%$ & $(58.17 \pm 0.92)\%$ & - & - & - & - & - & - \\
      $K_2^*(1430)^0$ & $(0.00 \pm 0.00)\%$ & $(-0.00 \pm 0.00)\%$ & $(1.68 \pm 0.29)\%$ & - & - & - & - & - \\
      $K_2^*(1430)^-$ & $(-0.86 \pm 0.23)\%$ & $(4.88 \pm 0.10)\%$ & $(1.23 \pm 0.11)\%$ & $(18.18 \pm 0.78)\%$ & - & - & - & - \\
      $K_3^*(1780)^0$ & $(-0.00 \pm 0.00)\%$ & $(0.00 \pm 0.00)\%$ & $(0.00 \pm 0.00)\%$ & $(0.71 \pm 0.11)\%$ & $(1.15 \pm 0.26)\%$ & - & - & - \\
      $K_4^*(2045)^-$ & $(0.10 \pm 0.04)\%$ & $(0.38 \pm 0.30)\%$ & $(0.12 \pm 0.03)\%$ & $(-0.00 \pm 0.00)\%$ & $(0.06 \pm 0.05)\%$ & $(0.42 \pm 0.14)\%$ & - & - \\
      $K^*(892)^-$ & $(-0.72 \pm 0.13)\%$ & $(0.73 \pm 0.32)\%$ & $(0.31 \pm 0.08)\%$ & $(0.00 \pm 0.00)\%$ & $(0.54 \pm 0.06)\%$ & $(-0.00 \pm 0.00)\%$ & $(5.33 \pm 0.41)\%$ & - \\
      $a_2(1320)^+$ & $(-0.36 \pm 0.11)\%$ & $(1.48 \pm 0.09)\%$ & $(-0.61 \pm 0.07)\%$ & $(1.96 \pm 0.14)\%$ & $(0.19 \pm 0.07)\%$ & $(0.17 \pm 0.06)\%$ & $(-0.24 \pm 0.08)\%$ & $(3.43 \pm 0.45)\%$ \\
      \hline
      Total & \multicolumn{1}{c}{$(90.57 \pm 1.50)\%$} & & & & & \\
      \bottomrule
      \end{tabular}}
  \label{tab:ff3080.0}
\end{table*}

\begin{table*}[htbp]\centering
  \caption{The nominal fit FFs (in units of \%) of each component and the interference for the dataset taken at $\sqrt s=3096.20$ MeV. Total refers to the sum of the FFs without considering interference. The uncertainties are statistical only.}
  \resizebox{\linewidth}{!}{
    \begin{tabular}{l|ccccccccccc}
      \toprule
      Mode & $NR_{K_S^0 K^+}(1^-)$ & $K^*(1680)^0$ & $K^*(892)^0$ & $K^*(1680)^-$ & $K_2^*(1430)^0$ & $K_2^*(1980)^0$ & $K_2^*(1430)^-$ & $K_2^*(1980)^-$ & $K^*(892)^-$ & $a_2(1320)^+$ & $\rho(1450)^+$ \\
      \hline
      $NR_{K_S^0 K^+}(1^-)$ & $(8.35 \pm 0.85)\%$ & - & - & - & - & - & - & - & - & - & - \\
      $K^*(1680)^0$ & $(-1.65 \pm 0.38)\%$ & $(0.43 \pm 0.13)\%$ & - & - & - & - & - & - & - & - & - \\
      $K^*(892)^0$ & $(9.08 \pm 0.39)\%$ & $(-0.04 \pm 0.20)\%$ & $(31.30 \pm 0.51)\%$ & - & - & - & - & - & - & - & - \\
      $K^*(1680)^-$ & $(-2.30 \pm 0.38)\%$ & $(0.24 \pm 0.04)\%$ & $(-1.10 \pm 0.14)\%$ & $(0.52 \pm 0.14)\%$ & - & - & - & - & - & - & - \\
      $K_2^*(1430)^0$ & $(-0.63 \pm 0.04)\%$ & $(-0.00 \pm 0.00)\%$ & $(-0.00 \pm 0.00)\%$ & $(-0.09 \pm 0.04)\%$ & $(1.50 \pm 0.15)\%$ & - & - & - & - & - & - \\
      $K_2^*(1980)^0$ & $(-0.40 \pm 0.06)\%$ & $(0.00 \pm 0.00)\%$ & $(-0.00 \pm 0.00)\%$ & $(-0.04 \pm 0.05)\%$ & $(0.26 \pm 0.08)\%$ & $(0.35 \pm 0.10)\%$ & - & - & - & - & - \\
      $K_2^*(1430)^-$ & $(-0.99 \pm 0.05)\%$ & $(-0.28 \pm 0.06)\%$ & $(-1.00 \pm 0.09)\%$ & $(-0.00 \pm 0.00)\%$ & $(0.56 \pm 0.03)\%$ & $(0.35 \pm 0.06)\%$ & $(4.09 \pm 0.28)\%$ & - & - & - & - \\
      $K_2^*(1980)^-$ & $(-0.52 \pm 0.06)\%$ & $(-0.18 \pm 0.05)\%$ & $(-0.87 \pm 0.15)\%$ & $(-0.00 \pm 0.00)\%$ & $(0.32 \pm 0.04)\%$ & $(0.22 \pm 0.03)\%$ & $(0.67 \pm 0.12)\%$ & $(0.58 \pm 0.13)\%$ & - & - & - \\
      $K^*(892)^-$ & $(11.07 \pm 0.45)\%$ & $(-1.02 \pm 0.18)\%$ & $(4.07 \pm 0.05)\%$ & $(-0.68 \pm 0.25)\%$ & $(-0.89 \pm 0.10)\%$ & $(-1.07 \pm 0.18)\%$ & $(0.00 \pm 0.00)\%$ & $(0.00 \pm 0.00)\%$ & $(43.06 \pm 0.63)\%$ & - & - \\
      $a_2(1320)^+$ & $(0.00 \pm 0.00)\%$ & $(-0.06 \pm 0.02)\%$ & $(-0.15 \pm 0.04)\%$ & $(0.03 \pm 0.02)\%$ & $(-0.14 \pm 0.03)\%$ & $(-0.10 \pm 0.03)\%$ & $(0.23 \pm 0.04)\%$ & $(0.14 \pm 0.03)\%$ & $(0.20 \pm 0.05)\%$ & $(0.20 \pm 0.07)\%$ & - \\
      $\rho(1450)^+$ & $(-2.88 \pm 0.30)\%$ & $(0.33 \pm 0.05)\%$ & $(-1.13 \pm 0.10)\%$ & $(0.36 \pm 0.06)\%$ & $(0.02 \pm 0.03)\%$ & $(-0.02 \pm 0.04)\%$ & $(0.05 \pm 0.05)\%$ & $(0.04 \pm 0.04)\%$ & $(-1.37 \pm 0.12)\%$ & $(-0.00 \pm 0.00)\%$ & $(0.94 \pm 0.15)\%$ \\
      \hline
      Total & \multicolumn{1}{c}{$(91.33 \pm 1.25)\%$} & & & & & \\
      \bottomrule
      \end{tabular}}
  \label{tab:ff3096.2}
\end{table*}

\section{OBSERVED CROSS SECTION MEASUREMENTS}

The observed cross section for $e^+e^-\to K_S^0K^+\pi^-$ at $\sqrt{s}$ is written as
\begin{equation}\label{cs}
  \sigma_{\rm obs}(\sqrt{s})=\frac{N_{\rm{sig}}}{\mathcal{L}\,\epsilon\,\mathcal{B}},
\end{equation}
where $N_\text{sig}$ is the number of signal events, and $\mathcal{L}$ is the integrated luminosity. The efficiency $\epsilon$ is obtained by weighting the MC simulation according to the PWA results. To account for the ISR and beam energy spread effects, the MC generation with ConExc is iteratively updated using the cross sections obtained from the line shape fit, until the change in the measured cross section between iterations becomes negligible (less than 0.5\%). The symbol $\mathcal{B}$ is the BF of $K^0_S\to \pi^+\pi^-$ quoted from the PDG~\cite{pdg}.

The observed cross section for the intermediate process, $e^+e^-\to\bar K^0 K^*(892)^0$ or $e^+e^-\to K^+ K^*(892)^-$, at each energy point is obtained with the same approach, where $N_\text{sig}$ is replaced with the product of the number of signal events for $e^+e^-\to K_S^0K^+\pi^-$ and the FFs according to the PWA results, $N_{{\rm sig},i}=N_{\rm sig}\mathrm{FF}_i^{\rm rec}$. The corresponding observed cross section is
\begin{equation}
  \sigma_{{\rm obs},i}(\sqrt{s})=\frac{N_{\rm sig}\mathrm{FF}_i^{\rm rec}}{\mathcal{L}\,\epsilon_i\,\mathcal{B}_i},
\end{equation}
where $\epsilon$ is obtained from signal MC samples of each specific process using its own cross section to iteration, and $\mathcal{B}$ is replaced with the product of the BF of $K_S^0\to\pi^+\pi^-$ and that of the intermediate states~[$\mathcal{B}(\bar{K}^0\to K_S^0)=50\%$, $\mathcal{B}(K^*(892)^0\to K^+\pi^-)=66.50\%$, $\mathcal{B}(K^*(892)^-\to \bar{K}^0\pi^-)=66.60\%$] from the PDG~\cite{pdg}. A correction factor is applied to account for data-MC discrepancies in the signal efficiencies, listed in Table~\ref{tab:corrfactor}. The observed cross sections of the processes $e^+e^-\to K_S^0K^+\pi^-$, $e^+e^-\to\bar K^0 K^*(892)^0$, and $e^+e^-\to K^+ K^*(892)^-$ are listed in Tables~\ref{tab:kskpixs}, \ref{tab:k0kstarxs}, and \ref{tab:kkstarxs}, respectively.

\begin{table}[htbp]\centering
  \caption{The used correction factors for different sources. The modes I, II and III refer to the decay processes of $J/\psi\to K_S^0K^+\pi^-$, $J/\psi\to \bar K^0 K^*(892)^0$ and $J/\psi\to K^+K^*(892)^-$, respectively.}
  \setlength{\tabcolsep}{0pt}
  \resizebox{\linewidth}{!}{
  \begin{tabular}{c@{\hspace{7pt}}c@{\hspace{5pt}}c@{\hspace{5pt}}c@{\hspace{7pt}}c@{\hspace{5pt}}c@{\hspace{5pt}}c@{\hspace{2pt}}}
  \toprule
  \multirow{2}{*}{$\sqrt{s}$ (\si{MeV})}   & \multicolumn{3}{c}{Kaon PID} & \multicolumn{3}{c}{Pion PID} \\
  \cmidrule(lr){2-7}
  & I & II & III & I & II & III \\
  \midrule 
  $3000.00 \pm 0.20$  & $1.0075$ & $1.0064$ & $1.0245$ & $0.9885$ & $0.9918$ & $0.9929$ \\
  $3020.00 \pm 0.20$  & $1.0081$ & $1.0067$ & $1.0252$ & $0.9882$ & $0.9917$ & $0.9926$ \\
  $3049.64 \pm 0.06$  & $1.0088$ & $1.0069$ & $1.0262$ & $0.9879$ & $0.9914$ & $0.9924$ \\
  $3058.69 \pm 0.06$  & $1.0088$ & $1.0070$ & $1.0266$ & $0.9889$ & $0.9913$ & $0.9925$ \\
  $3080.00 \pm 0.20$  & $1.0093$ & $1.0072$ & $1.0274$ & $0.9885$ & $0.9911$ & $0.9923$ \\
  $3082.50 \pm 0.06$  & $1.0106$ & $1.0073$ & $1.0275$ & $0.9880$ & $0.9911$ & $0.9922$ \\
  $3087.59 \pm 0.13$  & $1.0096$ & $1.0074$ & $1.0277$ & $0.9882$ & $0.9912$ & $0.9922$ \\
  $3088.85 \pm 0.06$  & $1.0108$ & $1.0074$ & $1.0277$ & $0.9898$ & $0.9911$ & $0.9922$ \\
  $3091.76 \pm 0.06$  & $1.0122$ & $1.0074$ & $1.0278$ & $0.9899$ & $0.9910$ & $0.9922$ \\
  $3094.70 \pm 0.10$  & $1.0164$ & $1.0075$ & $1.0279$ & $0.9894$ & $0.9910$ & $0.9921$ \\
  $3095.43 \pm 0.10$  & $1.0168$ & $1.0075$ & $1.0280$ & $0.9893$ & $0.9910$ & $0.9921$ \\
  $3095.73 \pm 0.08$  & $1.0170$ & $1.0074$ & $1.0280$ & $0.9892$ & $0.9910$ & $0.9921$ \\
  $3095.83 \pm 0.09$  & $1.0168$ & $1.0075$ & $1.0280$ & $0.9892$ & $0.9910$ & $0.9921$ \\
  $3096.20 \pm 0.07$  & $1.0172$ & $1.0075$ & $1.0280$ & $0.9892$ & $0.9911$ & $0.9921$ \\
  $3096.99 \pm 0.08$  & $1.0172$ & $1.0075$ & $1.0280$ & $0.9890$ & $0.9910$ & $0.9921$ \\
  $3097.21 \pm 0.09$  & $1.0170$ & $1.0075$ & $1.0281$ & $0.9890$ & $0.9910$ & $0.9920$ \\
  $3097.23 \pm 0.10$  & $1.0169$ & $1.0075$ & $1.0280$ & $0.9889$ & $0.9910$ & $0.9920$ \\
  $3097.65 \pm 0.08$  & $1.0172$ & $1.0075$ & $1.0281$ & $0.9891$ & $0.9910$ & $0.9920$ \\
  $3098.34 \pm 0.09$  & $1.0172$ & $1.0076$ & $1.0281$ & $0.9888$ & $0.9910$ & $0.9920$ \\
  $3098.73 \pm 0.08$  & $1.0171$ & $1.0075$ & $1.0281$ & $0.9888$ & $0.9909$ & $0.9920$ \\
  $3099.04 \pm 0.11$  & $1.0161$ & $1.0075$ & $1.0281$ & $0.9889$ & $0.9910$ & $0.9920$ \\
  $3101.36 \pm 0.12$  & $1.0169$ & $1.0076$ & $1.0280$ & $0.9892$ & $0.9910$ & $0.9921$ \\
  $3104.00 \pm 0.08$  & $1.0171$ & $1.0075$ & $1.0281$ & $0.9895$ & $0.9910$ & $0.9921$ \\
  $3105.58 \pm 0.10$  & $1.0169$ & $1.0075$ & $1.0280$ & $0.9897$ & $0.9910$ & $0.9920$ \\
  $3112.05 \pm 0.11$  & $1.0161$ & $1.0076$ & $1.0280$ & $0.9891$ & $0.9910$ & $0.9921$ \\
  $3119.88 \pm 0.13$  & $1.0136$ & $1.0076$ & $1.0280$ & $0.9882$ & $0.9910$ & $0.9920$ \\
  \bottomrule
  \label{tab:corrfactor}
  \end{tabular}}
\end{table}

\begin{table*}[htbp]\centering
  \caption{The observed cross sections of $e^+e^-\to K_S^0 K^+ \pi^-$ at different energy points. Here, $N_{\rm sig}$ is the observed signal yield; $\epsilon$ is the detection efficiency; $\mathcal{L}$ is the integrated luminosity; The uncertainties on $\sqrt{s}$ and $\mathcal{L}$ include both statistical and systematic contributions, while the uncertainties on the other numbers are statistical only.}
  \begin{tabular}{ccccc}
    \toprule
    $\sqrt{s}$ (\si{MeV}) & $N_{\rm sig}$ & $\epsilon$~(\%) & $\mathcal{L}$~(\si{pb^{-1}}) & $\sigma_{\rm obs}$ (\si{pb})\\
    \midrule
    $3000.00 \pm 0.20$ & $  272\pm  16$ & $37.00$ & $15.85 \pm 0.11$  & $   67.0\pm   4.1$ \\
    $3020.00 \pm 0.20$ & $  252\pm  16$ & $36.29$ & $17.32 \pm 0.12$  & $   58.0\pm   3.7$ \\
    $3049.64 \pm 0.06$ & $  238\pm  15$ & $37.11$ & $14.92 \pm 0.16$  & $   62.0\pm   4.0$ \\
    $3058.69 \pm 0.06$ & $  220\pm  15$ & $37.48$ & $15.06 \pm 0.16$  & $   56.3\pm   3.8$ \\
    $3080.00 \pm 0.20$ & $ 4381\pm  66$ & $35.36$ & $293.27 \pm 0.91$ & $   61.0\pm   0.9$ \\
    $3082.50 \pm 0.06$ & $   60\pm   8$ & $36.27$ & $4.77 \pm 0.06$   & $   50.1\pm   6.5$ \\
    $3087.59 \pm 0.13$ & $   36\pm   6$ & $36.02$ & $2.47 \pm 0.02$   & $   58.4\pm   9.7$ \\
    $3088.85 \pm 0.06$ & $  319\pm  18$ & $37.92$ & $15.56 \pm 0.17$  & $   78.3\pm   4.4$ \\
    $3091.76 \pm 0.06$ & $  377\pm  19$ & $38.61$ & $14.91 \pm 0.16$  & $   94.7\pm   4.9$ \\
    $3094.70 \pm 0.10$ & $  633\pm  25$ & $41.72$ & $2.14 \pm 0.03$   & $ 1022.7\pm  40.7$ \\
    $3095.43 \pm 0.10$ & $ 2033\pm  45$ & $41.62$ & $1.82 \pm 0.02$   & $ 3887.4\pm  86.2$ \\
    $3095.73 \pm 0.08$ & $ 5471\pm  74$ & $39.94$ & $2.92 \pm 0.02$   & $ 6779.0\pm  91.7$ \\
    $3095.83 \pm 0.09$ & $ 6084\pm  78$ & $41.72$ & $2.14 \pm 0.03$   & $ 9870.4\pm 126.5$ \\
    $3096.20 \pm 0.07$ & $17924\pm 134$ & $39.74$ & $4.98 \pm 0.03$   & $13089.1\pm  97.8$ \\
    $3096.99 \pm 0.08$ & $14717\pm 121$ & $39.48$ & $3.10 \pm 0.02$   & $17377.8\pm 143.2$ \\
    $3097.21 \pm 0.09$ & $10183\pm 101$ & $41.53$ & $2.07 \pm 0.03$   & $17123.7\pm 169.7$ \\
    $3097.23 \pm 0.10$ & $ 8083\pm  90$ & $39.88$ & $1.68 \pm 0.01$   & $17436.1\pm 193.9$ \\
    $3097.65 \pm 0.08$ & $17242\pm 131$ & $39.48$ & $4.66 \pm 0.03$   & $13542.3\pm 103.1$ \\
    $3098.34 \pm 0.09$ & $ 4933\pm  70$ & $41.29$ & $2.20 \pm 0.03$   & $ 7836.5\pm 111.6$ \\
    $3098.73 \pm 0.08$ & $ 7033\pm  84$ & $39.33$ & $5.64 \pm 0.03$   & $ 4581.5\pm  54.6$ \\
    $3099.04 \pm 0.11$ & $  638\pm  25$ & $41.22$ & $0.76 \pm 0.01$   & $ 2959.6\pm 117.1$ \\
    $3101.36 \pm 0.12$ & $  422\pm  21$ & $40.56$ & $1.61 \pm 0.02$   & $  932.4\pm  45.4$ \\
    $3104.00 \pm 0.08$ & $  878\pm  30$ & $38.86$ & $5.72 \pm 0.03$   & $  570.4\pm  19.3$ \\
    $3105.58 \pm 0.10$ & $  288\pm  17$ & $40.79$ & $2.11 \pm 0.03$   & $  484.1\pm  28.5$ \\
    $3112.05 \pm 0.11$ & $  150\pm  12$ & $39.88$ & $1.72 \pm 0.02$   & $  315.9\pm  25.8$ \\
    $3119.88 \pm 0.13$ & $   77\pm   9$ & $38.68$ & $1.26 \pm 0.02$   & $  228.5\pm  26.0$ \\
    \bottomrule
  \end{tabular}
  \label{tab:kskpixs}
\end{table*}

\begin{table*}[htbp]\centering
  \caption{The observed cross sections of $e^+e^-\to\bar K^0 K^*(892)^0$ at different energy points. Here, $N_{\rm sig}$ is the observed signal yield; $\epsilon$ is the detection efficiency; $\mathrm{FF}$ is the FF obtained with generator level PHSP MC events; $\mathrm{FF}^{\rm rec}$ is the FF obtained with reconstructed level PHSP MC events; $\mathcal{L}$ is the integrated luminosity; The uncertainties on $\sqrt{s}$ and $\mathcal{L}$ include both statistical and systematic contributions, while the uncertainties on the other numbers are statistical only.}
  \begin{tabular}{ccccccc}
    \toprule
    $\sqrt{s}$ (\si{MeV}) & $N_{\rm sig}$ & $\mathrm{FF}$~(\%) & $\mathrm{FF}^{\rm rec}$~(\%) & $\epsilon$~(\%) & $\mathcal{L}$~(\si{pb^{-1}}) & $\sigma_{\rm obs}$ (\si{pb})\\
    \midrule
    $3000.00 \pm 0.20$ & $ 167\pm  15$ & $56.88\pm  3.90$ & $61.31\pm  3.88$ & $40.52$ & $15.85 \pm 0.11$  & $  112.9\pm   9.9$ \\
    $3020.00 \pm 0.20$ & $ 145\pm  16$ & $52.46\pm  5.05$ & $57.25\pm  5.18$ & $40.40$ & $17.32 \pm 0.12$  & $   89.8\pm   9.9$ \\
    $3049.64 \pm 0.06$ & $ 137\pm  13$ & $52.58\pm  3.96$ & $57.53\pm  3.91$ & $41.27$ & $14.92 \pm 0.16$  & $   96.5\pm   9.1$ \\
    $3058.69 \pm 0.06$ & $ 143\pm  13$ & $60.56\pm  4.06$ & $65.23\pm  3.88$ & $41.63$ & $15.06 \pm 0.16$  & $   99.4\pm   8.9$ \\
    $3080.00 \pm 0.20$ & $2740\pm  57$ & $58.17\pm  0.92$ & $62.54\pm  0.88$ & $39.34$ & $293.27 \pm 0.91$ & $  103.2\pm   2.1$ \\
    $3082.50 \pm 0.06$ & $  28\pm   6$ & $42.59\pm  7.44$ & $47.37\pm  7.71$ & $41.38$ & $4.77 \pm 0.06$   & $   62.6\pm  13.0$ \\
    $3087.59 \pm 0.13$ & $  20\pm   6$ & $50.71\pm 12.19$ & $55.53\pm 12.43$ & $39.86$ & $2.47 \pm 0.02$   & $   88.2\pm  24.6$ \\
    $3088.85 \pm 0.06$ & $ 219\pm  16$ & $63.80\pm  3.42$ & $68.68\pm  3.26$ & $41.74$ & $15.56 \pm 0.17$  & $  146.8\pm  10.8$ \\
    $3091.76 \pm 0.06$ & $ 246\pm  20$ & $60.25\pm  3.91$ & $65.31\pm  3.98$ & $42.53$ & $14.91 \pm 0.16$  & $  169.0\pm  13.5$ \\
    $3094.70 \pm 0.10$ & $ 278\pm  25$ & $38.71\pm  3.22$ & $43.97\pm  3.52$ & $46.51$ & $2.14 \pm 0.03$   & $ 1213.1\pm 108.5$ \\
    $3095.43 \pm 0.10$ & $ 724\pm  35$ & $31.04\pm  1.41$ & $35.60\pm  1.55$ & $46.64$ & $1.82 \pm 0.02$   & $ 3714.0\pm 181.1$ \\
    $3095.73 \pm 0.08$ & $1907\pm  58$ & $30.39\pm  0.86$ & $34.86\pm  0.94$ & $45.09$ & $2.92 \pm 0.02$   & $ 6295.5\pm 190.3$ \\
    $3095.83 \pm 0.09$ & $2351\pm  71$ & $33.93\pm  0.97$ & $38.64\pm  1.06$ & $46.51$ & $2.14 \pm 0.03$   & $10288.2\pm 312.5$ \\
    $3096.20 \pm 0.07$ & $6403\pm 111$ & $31.30\pm  0.51$ & $35.72\pm  0.56$ & $44.81$ & $4.98 \pm 0.03$   & $12469.7\pm 216.7$ \\
    $3096.99 \pm 0.08$ & $4932\pm  92$ & $29.25\pm  0.51$ & $33.51\pm  0.56$ & $44.84$ & $3.10 \pm 0.02$   & $15419.6\pm 287.1$ \\
    $3097.21 \pm 0.09$ & $3560\pm  75$ & $30.41\pm  0.59$ & $34.96\pm  0.65$ & $46.79$ & $2.07 \pm 0.03$   & $15981.4\pm 337.3$ \\
    $3097.23 \pm 0.10$ & $2779\pm  69$ & $29.99\pm  0.70$ & $34.38\pm  0.77$ & $45.27$ & $1.68 \pm 0.01$   & $15879.6\pm 396.7$ \\
    $3097.65 \pm 0.08$ & $5874\pm 107$ & $29.73\pm  0.51$ & $34.07\pm  0.56$ & $44.70$ & $4.66 \pm 0.03$   & $12255.5\pm 223.5$ \\
    $3098.34 \pm 0.09$ & $1684\pm  54$ & $29.56\pm  0.88$ & $34.13\pm  0.97$ & $46.62$ & $2.20 \pm 0.03$   & $ 7125.3\pm 226.8$ \\
    $3098.73 \pm 0.08$ & $2286\pm  63$ & $28.32\pm  0.74$ & $32.51\pm  0.81$ & $44.87$ & $5.64 \pm 0.03$   & $ 3926.8\pm 108.7$ \\
    $3099.04 \pm 0.11$ & $ 260\pm  20$ & $35.68\pm  2.57$ & $40.66\pm  2.77$ & $46.17$ & $0.76 \pm 0.01$   & $ 3231.9\pm 254.5$ \\
    $3101.36 \pm 0.12$ & $ 128\pm  15$ & $26.86\pm  2.88$ & $30.28\pm  3.14$ & $45.99$ & $1.61 \pm 0.02$   & $  749.1\pm  85.8$ \\
    $3104.00 \pm 0.08$ & $ 329\pm  24$ & $31.93\pm  2.22$ & $37.49\pm  2.48$ & $43.65$ & $5.72 \pm 0.03$   & $  572.7\pm  42.5$ \\
    $3105.58 \pm 0.10$ & $  96\pm  12$ & $28.95\pm  3.47$ & $33.36\pm  3.82$ & $45.30$ & $2.11 \pm 0.03$   & $  437.4\pm  56.3$ \\
    $3112.05 \pm 0.11$ & $  45\pm   8$ & $25.53\pm  4.21$ & $30.34\pm  4.72$ & $44.21$ & $1.72 \pm 0.02$   & $  260.0\pm  45.6$ \\
    $3119.88 \pm 0.13$ & $  35\pm   7$ & $39.80\pm  6.96$ & $45.09\pm  7.35$ & $42.80$ & $1.26 \pm 0.02$   & $  280.1\pm  55.7$ \\
    \bottomrule
  \end{tabular}
  \label{tab:k0kstarxs}
\end{table*}

\begin{table*}[htbp]\centering
  \caption{The observed cross sections of $e^+e^-\to K^+ K^*(892)^-$ at different energy points. Here, $N_{\rm sig}$ is the observed signal yield; $\epsilon$ is the detection efficiency; $\mathrm{FF}$ is the FF obtained with generator level PHSP MC events; $\mathrm{FF}^{\rm rec}$ is the FF obtained with reconstructed level PHSP MC events; $\mathcal{L}$ is the integrated luminosity; The uncertainties on $\sqrt{s}$ and $\mathcal{L}$ include both statistical and systematic contributions, while the uncertainties on the other numbers are statistical only.}
  \begin{tabular}{ccccccc}
    \toprule
    $\sqrt{s}$ (\si{MeV}) & $N_{\rm sig}$ & $\mathrm{FF}$~(\%) & $\mathrm{FF}^{\rm rec}$~(\%) & $\epsilon$~(\%) & $\mathcal{L}$~(\si{pb^{-1}}) & $\sigma_{\rm obs}$ (\si{pb})\\
    \midrule
    $3000.00 \pm 0.20$ & $   2\pm   2$ & $ 0.67\pm  0.73$ & $ 0.59\pm  0.64$ & $34.94$ & $15.85 \pm 0.11$  & $    1.3\pm   1.4$ \\
    $3020.00 \pm 0.20$ & $   2\pm   3$ & $ 0.81\pm  1.21$ & $ 0.73\pm  1.09$ & $34.86$ & $17.32 \pm 0.12$  & $    1.3\pm   2.0$ \\
    $3049.64 \pm 0.06$ & $   7\pm   3$ & $ 3.52\pm  1.60$ & $ 3.09\pm  1.42$ & $36.36$ & $14.92 \pm 0.16$  & $    5.9\pm   2.7$ \\
    $3058.69 \pm 0.06$ & $   4\pm   2$ & $ 1.94\pm  1.21$ & $ 1.73\pm  1.08$ & $37.13$ & $15.06 \pm 0.16$  & $    2.9\pm   1.8$ \\
    $3080.00 \pm 0.20$ & $ 205\pm  16$ & $ 5.33\pm  0.41$ & $ 4.67\pm  0.36$ & $36.05$ & $293.27 \pm 0.91$ & $    8.4\pm   0.7$ \\
    $3082.50 \pm 0.06$ & $   3\pm   2$ & $ 5.37\pm  3.31$ & $ 4.97\pm  3.07$ & $38.34$ & $4.77 \pm 0.06$   & $    7.1\pm   4.5$ \\
    $3087.59 \pm 0.13$ & $   4\pm   2$ & $11.78\pm  6.25$ & $10.50\pm  5.64$ & $36.76$ & $2.47 \pm 0.02$   & $   18.0\pm  10.2$ \\
    $3088.85 \pm 0.06$ & $  38\pm   7$ & $13.81\pm  2.27$ & $11.94\pm  1.99$ & $38.82$ & $15.56 \pm 0.17$  & $   27.4\pm   4.8$ \\
    $3091.76 \pm 0.06$ & $  85\pm  11$ & $25.91\pm  3.00$ & $22.60\pm  2.68$ & $39.31$ & $14.91 \pm 0.16$  & $   63.2\pm   8.2$ \\
    $3094.70 \pm 0.10$ & $ 235\pm  19$ & $40.17\pm  2.72$ & $37.16\pm  2.58$ & $39.68$ & $2.14 \pm 0.03$   & $ 1200.0\pm  96.1$ \\
    $3095.43 \pm 0.10$ & $ 752\pm  39$ & $39.86\pm  1.81$ & $37.01\pm  1.71$ & $39.79$ & $1.82 \pm 0.02$   & $ 4519.2\pm 231.6$ \\
    $3095.73 \pm 0.08$ & $2202\pm  67$ & $43.39\pm  1.17$ & $40.25\pm  1.11$ & $37.91$ & $2.92 \pm 0.02$   & $ 8632.9\pm 264.4$ \\
    $3095.83 \pm 0.09$ & $2355\pm  63$ & $41.70\pm  0.95$ & $38.71\pm  0.90$ & $39.66$ & $2.14 \pm 0.03$   & $12071.2\pm 321.4$ \\
    $3096.20 \pm 0.07$ & $7210\pm 119$ & $43.06\pm  0.63$ & $40.23\pm  0.59$ & $37.51$ & $4.98 \pm 0.03$   & $16752.4\pm 277.4$ \\
    $3096.99 \pm 0.08$ & $5949\pm 115$ & $43.14\pm  0.74$ & $40.42\pm  0.71$ & $37.51$ & $3.10 \pm 0.02$   & $22204.7\pm 429.5$ \\
    $3097.21 \pm 0.09$ & $4008\pm  86$ & $42.21\pm  0.79$ & $39.36\pm  0.75$ & $39.61$ & $2.07 \pm 0.03$   & $21220.8\pm 454.8$ \\
    $3097.23 \pm 0.10$ & $3303\pm  77$ & $43.78\pm  0.88$ & $40.86\pm  0.84$ & $37.72$ & $1.68 \pm 0.01$   & $22618.3\pm 529.6$ \\
    $3097.65 \pm 0.08$ & $6966\pm 122$ & $43.12\pm  0.67$ & $40.40\pm  0.64$ & $37.43$ & $4.66 \pm 0.03$   & $17330.0\pm 304.4$ \\
    $3098.34 \pm 0.09$ & $1950\pm  60$ & $42.37\pm  1.14$ & $39.53\pm  1.09$ & $39.81$ & $2.20 \pm 0.03$   & $ 9648.3\pm 299.2$ \\
    $3098.73 \pm 0.08$ & $2817\pm  83$ & $42.68\pm  1.13$ & $40.05\pm  1.08$ & $37.69$ & $5.64 \pm 0.03$   & $ 5750.1\pm 169.2$ \\
    $3099.04 \pm 0.11$ & $ 226\pm  19$ & $37.99\pm  2.73$ & $35.34\pm  2.60$ & $39.23$ & $0.76 \pm 0.01$   & $ 3300.2\pm 275.6$ \\
    $3101.36 \pm 0.12$ & $ 140\pm  16$ & $35.86\pm  3.62$ & $33.18\pm  3.45$ & $39.58$ & $1.61 \pm 0.02$   & $  952.2\pm 109.4$ \\
    $3104.00 \pm 0.08$ & $ 347\pm  24$ & $43.40\pm  2.51$ & $39.51\pm  2.37$ & $37.06$ & $5.72 \pm 0.03$   & $  709.7\pm  48.8$ \\
    $3105.58 \pm 0.10$ & $  99\pm  12$ & $37.51\pm  3.87$ & $34.43\pm  3.64$ & $39.14$ & $2.11 \pm 0.03$   & $  521.6\pm  63.1$ \\
    $3112.05 \pm 0.11$ & $  36\pm   8$ & $26.65\pm  5.49$ & $24.05\pm  5.03$ & $38.70$ & $1.72 \pm 0.02$   & $  235.1\pm  52.8$ \\
    $3119.88 \pm 0.13$ & $  18\pm   6$ & $24.63\pm  7.45$ & $22.67\pm  6.90$ & $37.21$ & $1.26 \pm 0.02$   & $  161.7\pm  52.5$ \\
    \bottomrule
  \end{tabular}
  \label{tab:kkstarxs}
\end{table*}

\section{SYSTEMATIC UNCERTAINTIES FOR CROSS SECTION MEASUREMENTS}\label{sec:sys}

The systematic uncertainties for the cross section measurements are categorized into two parts. The first arises from tracking, PID, $K_S^0$ reconstruction, luminosity measurement, kinematic fit, radiation, and vacuum polarization (VP) correction and from the quoted input BFs, all of which are unrelated to the PWA, and they are summarized in Table~\ref{tab:sysxsunrelate}. The second part is PWA related, which is summarized in Table~\ref{tab:totalpways}.

\begin{table*}[htbp]\centering
  \caption{The systematic uncertainties (in \%) of observed cross sections of $e^+e^-\to K_S^0 K^+\pi^-$ and all subprocesses unrelated to PWA. The modes I, II, and III refer to the decay processes of $J/\psi\to K_S^0K^+\pi^-$, $J/\psi\to \bar K^0 K^*(892)^0$, and $J/\psi\to K^+K^*(892)^-$, respectively.}
  \setlength{\tabcolsep}{0pt}
  \begin{tabular}{@{}
    c@{\hspace{5pt}}
    c@{\hspace{5pt}}
    c@{\hspace{5pt}}
    c@{\hspace{5pt}}
    c@{\hspace{5pt}}
    c@{\hspace{5pt}}
    c@{\hspace{5pt}}
    c@{\hspace{5pt}}
    c@{\hspace{5pt}}
    c@{\hspace{5pt}}
    c
    @{}}
    \toprule
    \multirow{2}{*}{$\sqrt{s}$ (\si{MeV}) }       &
    \multirow{2}{*}{Tracking}           &
    \multirow{2}{*}{PID} &
    \multirow{2}{*}{$K_S^0$ rec} &
    \multirow{2}{*}{Lumi} &
    \multirow{2}{*}{VP} &
    \multirow{2}{*}{Kinematic} &
    \multirow{2}{*}{Quoted BF} &
    \multicolumn{3}{c}{ISR}            \\
    \cmidrule(lr){9-11}
    & & & & & & & & I & II & III   \\
    \midrule
    $3000.00 \pm 0.20$ & 2.0 & 0.4 & 1.5 & 0.7 & 0.5 & 0.5 & 0.1 & 0.3 & 0.5 & 0.5 \\
    $3020.00 \pm 0.20$ & 2.0 & 0.4 & 1.5 & 0.7 & 0.5 & 0.5 & 0.1 & 0.1 & 0.8 & 0.3 \\
    $3049.64 \pm 0.06$ & 2.0 & 0.4 & 1.5 & 1.1 & 0.5 & 0.5 & 0.1 & 0.5 & 0.2 & 0.1 \\
    $3058.69 \pm 0.06$ & 2.0 & 0.4 & 1.5 & 1.1 & 0.5 & 0.5 & 0.1 & 0.1 & 0.7 & 0.5 \\
    $3080.00 \pm 0.20$ & 2.0 & 0.4 & 1.5 & 0.3 & 0.5 & 0.5 & 0.1 & 1.1 & 0.4 & 1.5 \\
    $3082.50 \pm 0.06$ & 2.0 & 0.4 & 1.5 & 1.3 & 0.5 & 0.5 & 0.1 & 1.5 & 0.3 & 0.1 \\
    $3087.59 \pm 0.13$ & 2.0 & 0.4 & 1.5 & 0.8 & 0.5 & 0.5 & 0.1 & 1.7 & 0.3 & 0.5 \\
    $3088.85 \pm 0.06$ & 2.0 & 0.4 & 1.5 & 1.1 & 0.5 & 0.5 & 0.1 & 1.5 & 0.1 & 1.1 \\
    $3091.76 \pm 0.06$ & 2.0 & 0.4 & 1.5 & 1.1 & 0.5 & 0.5 & 0.1 & 2.5 & 0.4 & 1.4 \\
    $3094.70 \pm 0.10$ & 2.0 & 0.4 & 1.5 & 1.4 & 0.5 & 0.5 & 0.1 & 0.2 & 0.2 & 0.5 \\
    $3095.43 \pm 0.10$ & 2.0 & 0.4 & 1.5 & 1.1 & 0.5 & 0.5 & 0.1 & 0.1 & 0.1 & 0.3 \\
    $3095.73 \pm 0.08$ & 2.0 & 0.4 & 1.5 & 0.7 & 0.5 & 0.5 & 0.1 & 0.1 & 0.1 & 0.1 \\
    $3095.83 \pm 0.09$ & 2.0 & 0.4 & 1.5 & 1.4 & 0.5 & 0.5 & 0.1 & 0.1 & 0.5 & 0.1 \\
    $3096.20 \pm 0.07$ & 2.0 & 0.4 & 1.5 & 0.6 & 0.5 & 0.5 & 0.1 & 0.5 & 0.4 & 0.1 \\
    $3096.99 \pm 0.08$ & 2.0 & 0.4 & 1.5 & 0.6 & 0.5 & 0.5 & 0.1 & 0.4 & 0.1 & 0.5 \\
    $3097.21 \pm 0.09$ & 2.0 & 0.4 & 1.5 & 1.4 & 0.5 & 0.5 & 0.1 & 0.2 & 0.3 & 0.1 \\
    $3097.23 \pm 0.10$ & 2.0 & 0.4 & 1.5 & 0.6 & 0.5 & 0.5 & 0.1 & 0.3 & 0.2 & 0.6 \\
    $3097.65 \pm 0.08$ & 2.0 & 0.4 & 1.5 & 0.6 & 0.5 & 0.5 & 0.1 & 0.3 & 0.3 & 0.6 \\
    $3098.34 \pm 0.09$ & 2.0 & 0.4 & 1.5 & 1.4 & 0.5 & 0.5 & 0.1 & 1.0 & 0.4 & 0.3 \\
    $3098.73 \pm 0.08$ & 2.0 & 0.4 & 1.5 & 0.5 & 0.5 & 0.5 & 0.1 & 0.4 & 0.1 & 0.6 \\
    $3099.04 \pm 0.11$ & 2.0 & 0.4 & 1.5 & 1.3 & 0.5 & 0.5 & 0.1 & 0.6 & 0.1 & 0.2 \\
    $3101.36 \pm 0.12$ & 2.0 & 0.4 & 1.5 & 1.2 & 0.5 & 0.5 & 0.1 & 0.1 & 0.2 & 0.1 \\
    $3104.00 \pm 0.08$ & 2.0 & 0.4 & 1.5 & 0.5 & 0.5 & 0.5 & 0.1 & 0.8 & 0.3 & 0.5 \\
    $3105.58 \pm 0.10$ & 2.0 & 0.4 & 1.5 & 1.4 & 0.5 & 0.5 & 0.1 & 0.4 & 0.4 & 0.1 \\
    $3112.05 \pm 0.11$ & 2.0 & 0.4 & 1.5 & 1.2 & 0.5 & 0.5 & 0.1 & 0.5 & 1.1 & 0.2 \\
    $3119.88 \pm 0.13$ & 2.0 & 0.4 & 1.5 & 1.6 & 0.5 & 0.5 & 0.1 & 0.1 & 0.4 & 0.7 \\
    \bottomrule
  \end{tabular}
  \label{tab:sysxsunrelate}
\end{table*}

The uncertainties due to kaon and pion tracking are studied with the control samples of $J/\psi\to \pi^+\pi^-\pi^0$ and $\psi(2S)\to \pi^+\pi^-J/\psi(J/\psi\to K^+K^-K^+K^-)$, the data MC differences are studied within the two-dimensional intervals of polar angle or azimuthal angle and transverse momentum, and the discrepancies between data and MC are estimated to be less than 1.0\%. Consequently, the tracking efficiency uncertainties are estimated to be 1.0\% per charged track~\cite{pionsys,kaonsys}, giving 2.0\% in total. The tracking of the $\pi^+\pi^-$ pair from the $K_S^0$ decay is included in the uncertainty of $K_S^0$ reconstruction. The systematic uncertainty of kaon PID is studied with the control samples of $e^+e^-\to K^+K^-K^+K^-$ and $e^+e^-\to K^+K^-\pi^+\pi^-(\pi^0)$, while systematic uncertainty of pion PID is studied with the control samples of $e^+e^-\to \pi^+\pi^-\pi^+\pi^-(\pi^0)$ and $e^+e^-\to K^+K^-\pi^+\pi^-(\pi^0)$ collected at $\sqrt{s} = 3.773$ GeV~\cite{3773data}. The differences of PID efficiencies between data and MC simulation are obtained in different two-dimensional intervals of polar angle and momentum. After weighting them according to the distribution of the polar angle and momentum of kaon and pion in the signal decay, we correct the MC efficiency and the residual uncertainty, $0.2\%$, is taken as the systematic uncertainty of both kaon and pion PID, as listed in Table~\ref{tab:corrfactor}. The efficiency of the $K_S^0$ reconstruction incorporating the tracking and the mass window selection is studied using the control samples of $J/\psi \to K^*(892)^\pm K^\mp$, with $K^*(892)^\pm\to K_S^0\pi^\pm$ and $J/\psi\to \phi K_S^0 K^\pm \pi^\mp$. The difference of $K_S^0$ reconstruction efficiency between data and MC simulation, $1.5\%$, is taken as the systematic uncertainty~\cite{kssys}. The uncertainty associated with the integrated luminosity is around 1\% at each energy point~\cite{wangyadi,jlumi,jpsinumber}. The uncertainty related to the kinematic fit is estimated by correcting the helix parameters of the simulated charged tracks to match the resolution~\cite{helix}. The uncertainty associated with the ISR and VP is obtained with the accuracy of the radiation function, which is about 0.5\%~\cite{ISRfactor}. The systematic uncertainties of the BFs of intermediate states in the subsequent decays are taken from the PDG~\cite{pdg}.

The second category of uncertainties is associated with the PWA fit. Fits with alternative scenarios are performed, and the changes of FFs are taken as the systematic uncertainties (see Table~\ref{tab:totalpways}). 

The systematic uncertainties associated with the resonance parameters, which are taken from the PDG and fixed in the fit, are estimated by performing alternative fits with the parameters varied by $\pm 1\sigma$. The systematic uncertainties associated with the additional resonances are estimated by alternative fits including the components $K_3^*(1780)^-$ or the $\rho_3(1690)^+$, which are the most significant, even if with a significance less than $5 \sigma$ obtained from data. To estimate the potential background contribution, we reintroduce the background using the data events in the $K_S^0$ sideband region and take the difference as the systematic uncertainty. The systematic uncertainties due to the barrier factor are estimated by varying the radius of the centrifugal barrier by $\pm 1\sigma$~\cite{BESIIIlambdapipi0}. In the nominal fit, the Breit-Wigner widths are treated as energy-dependent running widths; to evaluate the systematic uncertainty associated with this dynamic model, an alternative fit is performed in which the widths are treated as the constant values. The differences between the two fits are taken as the systematic uncertainties. The systematic uncertainty from MC modeling is estimated using an alternative signal MC sample, in which all fitted complex coupling parameters are randomly generated according to a multivariate Gaussian distribution based on the covariance matrix of the nominal fit. The difference in detection efficiencies between the alternative and nominal MC samples is taken as the corresponding uncertainty. In addition, 300 pseudoexperiments are performed using the same alternative MC sample to verify the stability of the fit. The extracted FFs are consistent with the input values, indicating no fit bias.

Assuming all the sources of systematic uncertainties as independent, the total uncertainties are the quadratic sums of the individual values.

\begin{table*}[htbp]\centering
  \caption{The systematic uncertainties (in \%) for various observables. The modes I, II, and III refer to the decay processes of $J/\psi\to K_S^0K^+\pi^-$, $J/\psi\to \bar K^0 K^*(892)^0$, and $J/\psi\to K^+K^*(892)^-$, respectively.}
  \setlength{\tabcolsep}{0pt}
  \resizebox{\linewidth}{!}{
    \begin{tabular}{@{}
      c@{\hspace{3pt}}
      c@{\hspace{3pt}}
      c@{\hspace{9pt}}c@{\hspace{9pt}}c@{\hspace{9pt}}
      c@{\hspace{9pt}}c@{\hspace{9pt}}c@{\hspace{9pt}}
      c@{\hspace{9pt}}c@{\hspace{9pt}}c@{\hspace{9pt}}
      c@{\hspace{9pt}}c@{\hspace{9pt}}c@{\hspace{9pt}}
      c@{\hspace{9pt}}c@{\hspace{9pt}}c@{\hspace{9pt}}
      c@{\hspace{9pt}}c@{\hspace{9pt}}c@{\hspace{9pt}}
      @{}}
      \toprule
      \multirow{2}{*}{$\sqrt{s}$ (\si{MeV}) }  &
      \multirow{2}{*}{MC model}                &
      \multicolumn{3}{c}{BKG description}      &
      \multicolumn{3}{c}{$M$ or $\Gamma$}      &
      \multicolumn{3}{c}{Extra resonance}      &
      \multicolumn{3}{c}{Radius}               &
      \multicolumn{3}{c}{Kinematic model}      &
      \multicolumn{3}{c}{Total}                \\
      \cmidrule(lr){3-5} \cmidrule(lr){6-8} \cmidrule(lr){9-11} \cmidrule(lr){12-14} \cmidrule(lr){15-17} \cmidrule(lr){18-20}
      & & I & II & III   &
          I & II & III   &
          I & II & III   &
          I & II & III   &
          I & II & III   &
          I & II & III   \\
      \midrule
      $3000.00 \pm 0.20$ & 1.1  & 0.1 & 0.4 & 0.5  & 0.1 & 0.7 & 1.9  & 0.0 &  0.3 & 25.4  & 0.2 & 0.5 & 0.5  & 0.1 & 1.5 &  0.1  & 1.1 &  1.8 & 25.5 \\
      $3020.00 \pm 0.20$ & 1.1  & 0.1 & 0.4 & 0.5  & 0.1 & 0.7 & 1.8  & 0.0 &  2.1 & 17.3  & 0.2 & 1.0 & 3.6  & 0.0 & 0.1 & 16.3  & 1.1 &  2.5 & 24.1 \\
      $3049.64 \pm 0.06$ & 1.1  & 0.1 & 0.4 & 0.5  & 0.1 & 0.7 & 1.8  & 0.2 &  0.3 &  1.4  & 0.3 & 0.3 & 0.7  & 0.0 & 0.7 &  1.4  & 1.1 &  1.2 &  2.8 \\
      $3058.69 \pm 0.06$ & 1.1  & 0.1 & 0.4 & 0.5  & 0.1 & 0.6 & 1.8  & 0.6 &  1.1 &  7.5  & 0.3 & 0.6 & 5.4  & 0.1 & 0.5 &  4.5  & 1.3 &  1.5 & 10.4 \\
      $3080.00 \pm 0.20$ & 0.2  & 0.1 & 0.4 & 0.5  & 0.1 & 0.6 & 1.8  & 0.0 &  0.6 &  0.3  & 0.3 & 0.1 & 0.5  & 0.1 & 0.3 &  1.2  & 0.4 &  1.0 &  2.3 \\
      $3082.50 \pm 0.06$ & 2.8  & 0.1 & 0.4 & 0.5  & 0.2 & 0.9 & 1.8  & 0.7 &  2.4 &  0.1  & 0.3 & 0.1 & 0.5  & 0.0 & 0.9 &  3.7  & 2.9 &  2.8 &  4.2 \\
      $3087.59 \pm 0.13$ & 3.7  & 0.1 & 0.4 & 0.5  & 0.1 & 0.6 & 1.6  & 0.8 &  1.3 &  0.2  & 0.4 & 0.1 & 0.8  & 0.3 & 1.7 &  1.1  & 3.9 &  2.3 &  2.2 \\
      $3088.85 \pm 0.06$ & 1.0  & 0.1 & 0.4 & 0.5  & 0.1 & 0.5 & 1.7  & 0.0 &  0.1 &  0.3  & 0.2 & 0.0 & 1.8  & 0.0 & 0.9 &  1.6  & 1.0 &  1.1 &  3.0 \\
      $3091.76 \pm 0.06$ & 2.0  & 0.1 & 0.4 & 0.5  & 0.1 & 0.5 & 1.5  & 0.2 &  0.3 &  0.8  & 0.1 & 1.9 & 1.2  & 0.0 & 1.3 &  1.9  & 2.0 &  2.4 &  2.9 \\
      $3094.70 \pm 0.10$ & 0.6  & 0.0 & 0.4 & 0.2  & 0.1 & 0.9 & 1.2  & 0.1 &  0.5 &  0.1  & 0.0 & 3.0 & 7.0  & 0.0 & 9.3 &  8.3  & 0.6 &  9.8 & 10.9 \\
      $3095.43 \pm 0.10$ & 0.3  & 0.0 & 0.4 & 0.2  & 0.1 & 1.0 & 1.1  & 0.0 &  0.1 &  0.1  & 0.1 & 4.3 & 3.0  & 0.0 & 1.0 &  1.4  & 0.3 &  4.6 &  3.5 \\
      $3095.73 \pm 0.08$ & 0.2  & 0.0 & 0.4 & 0.2  & 0.1 & 1.0 & 1.0  & 0.0 &  1.1 &  0.0  & 0.1 & 2.6 & 1.8  & 0.0 & 0.5 &  0.1  & 0.2 &  3.1 &  2.1 \\
      $3095.83 \pm 0.09$ & 0.2  & 0.0 & 0.4 & 0.2  & 0.1 & 1.0 & 1.1  & 0.0 &  0.1 &  0.1  & 0.1 & 0.3 & 1.6  & 0.0 & 0.9 &  0.4  & 0.2 &  1.4 &  2.0 \\
      $3096.20 \pm 0.07$ & 0.1  & 0.0 & 0.4 & 0.2  & 0.1 & 1.0 & 1.1  & 0.0 &  0.7 &  0.0  & 0.1 & 2.8 & 2.1  & 0.0 & 0.7 &  0.3  & 0.2 &  3.1 &  2.4 \\
      $3096.99 \pm 0.08$ & 0.1  & 0.0 & 0.4 & 0.2  & 0.1 & 1.0 & 1.1  & 0.0 &  0.2 &  0.2  & 0.1 & 3.7 & 1.8  & 0.0 & 0.9 &  0.5  & 0.2 &  4.0 &  2.2 \\
      $3097.21 \pm 0.09$ & 0.1  & 0.0 & 0.4 & 0.2  & 0.1 & 1.0 & 1.1  & 0.0 &  0.4 &  0.1  & 0.1 & 3.3 & 2.2  & 0.1 & 0.3 &  0.2  & 0.2 &  3.5 &  2.5 \\
      $3097.23 \pm 0.10$ & 0.1  & 0.0 & 0.4 & 0.2  & 0.1 & 1.0 & 1.0  & 0.0 &  0.3 &  0.1  & 0.0 & 3.6 & 2.3  & 0.1 & 0.7 &  0.2  & 0.2 &  3.9 &  2.5 \\
      $3097.65 \pm 0.08$ & 0.1  & 0.0 & 0.4 & 0.2  & 0.1 & 1.0 & 1.1  & 0.0 &  0.3 &  0.1  & 0.1 & 3.2 & 1.6  & 0.0 & 0.8 &  0.8  & 0.1 &  3.5 &  2.1 \\
      $3098.34 \pm 0.09$ & 0.2  & 0.0 & 0.4 & 0.2  & 0.1 & 1.0 & 1.1  & 0.0 &  0.1 &  0.2  & 0.1 & 3.0 & 1.7  & 0.0 & 0.6 &  0.5  & 0.2 &  3.2 &  2.0 \\
      $3098.73 \pm 0.08$ & 0.2  & 0.0 & 0.4 & 0.2  & 0.1 & 1.0 & 1.1  & 0.0 &  1.8 &  0.1  & 0.1 & 4.1 & 1.5  & 0.0 & 1.3 &  1.4  & 0.2 &  4.8 &  2.3 \\
      $3099.04 \pm 0.11$ & 0.6  & 0.0 & 0.4 & 0.2  & 0.1 & 0.9 & 1.2  & 0.0 &  2.8 &  4.8  & 0.0 & 0.8 & 1.1  & 0.0 & 1.1 &  1.5  & 0.6 &  3.3 &  5.3 \\
      $3101.36 \pm 0.12$ & 0.6  & 0.0 & 0.4 & 0.2  & 0.1 & 1.0 & 1.1  & 0.4 & 15.3 &  5.2  & 0.0 & 7.8 & 4.5  & 0.2 & 6.1 &  1.8  & 0.8 & 18.3 &  7.2 \\
      $3104.00 \pm 0.08$ & 0.5  & 0.0 & 0.4 & 0.2  & 0.1 & 1.0 & 1.1  & 0.0 &  0.1 &  0.1  & 0.1 & 1.3 & 0.4  & 0.0 & 0.3 &  0.8  & 0.5 &  1.7 &  1.4 \\
      $3105.58 \pm 0.10$ & 0.8  & 0.0 & 0.4 & 0.2  & 0.1 & 0.9 & 1.2  & 0.0 &  4.2 &  0.7  & 0.2 & 5.4 & 5.5  & 0.1 & 0.3 &  1.1  & 0.9 &  6.9 &  5.7 \\
      $3112.05 \pm 0.11$ & 1.1  & 0.0 & 0.4 & 0.2  & 0.1 & 0.9 & 1.3  & 0.2 &  5.5 &  0.7  & 0.3 & 7.1 & 7.3  & 0.0 & 0.1 &  1.7  & 1.2 &  9.0 &  7.6 \\
      $3119.88 \pm 0.13$ & 2.6  & 0.0 & 0.4 & 0.2  & 0.1 & 0.7 & 1.4  & 0.0 &  1.9 &  0.2  & 0.0 & 2.4 & 1.7  & 0.3 & 0.7 &  6.3  & 2.6 &  3.2 &  6.7 \\
      \bottomrule
  \end{tabular}}
  \label{tab:totalpways}
\end{table*}

\section{FIT TO THE LINE SHAPE}\label{sec:xsfit}

In order to obtain the desired physical parameters--the relative magnitudes and phases between strong and EM amplitude as well as BFs, a $\chi^2$ fit accounting for the correlated and uncorrelated uncertainties across different energy points is performed. The fit PDF is parametrized as the coherent sum of the continuum and resonant amplitudes, as described in Eq.~\eqref{eq:xsformular}. Because of the narrow width of the $J/\psi$ resonance, the observed cross section is strongly affected by ISR and the beam energy spread. To properly incorporate these effects, the fit is performed directly to the observed cross section.

Considering the VP effect, the dressed cross section is
\begin{equation}
  \sigma_\text{Dress}^{f}=\frac{\sigma_\text{Born}^{f}}{\left|1-\Pi_0(s)\right|^2},
\end{equation}
where $\Pi_0$ is the nonresonant part of energy dependence of the full VP~\cite{VPfactor} operator $\Pi$. The contribution from $J/\psi$ resonance production is included by the Breit-Wigner function with definitions of $\Gamma_{ee}=\Gamma_{ee}^0 /\left|1-\Pi_0\right|^2$ and $\Gamma_{\mu\mu}=\Gamma_{\mu\mu}^0 /\left|1-\Pi_0\right|^2$;
the $\Gamma_{ee}$ and $\Gamma_{\mu\mu}$ are the so-called experimental partial widths quoted from the PDG~\cite{pdg}.

The ISR effect is considered by
\begin{equation}
  \sigma^{\prime}(s)=\int_0^{1-\frac{s_\text{min}}{s}}dx F_{\rm ISR}(x, s)\sigma_\text{Dress}(s(1-x)),
\end{equation}
where $s_\text{min}=(2.85\,\text{GeV})^2$ represents the threshold of the cross section in the ISR integral, $x$ is the fraction of the c.m. energy lost through ISR photon emission, and $F_{\rm ISR}(x, s)$ is the radiative function by Kuraev and Fadin~\cite{ISRfactor} up to $\mathcal{O}(\alpha^2)$~\cite{conexc}:
\begin{widetext}
\begin{equation}
  F_{\rm ISR}(x, s)=x^{\beta-1} \beta (1+\delta)-\beta(1-\frac{x}{2})+\frac{\beta^2}{8}\left[4(2-x) \ln \frac{1}{x}-\frac{1+3(1-x)^2}{x} \ln (1-x)-6+x\right],
\end{equation}
\end{widetext}
with $\delta=\frac{3}{4} \beta+\frac{\alpha}{\pi}\left(\frac{\pi^2}{3}-\frac{1}{2}\right)+\beta^2\left(\frac{9}{32}-\frac{\pi^2}{12}\right)$, $\beta=\frac{2 \alpha}{\pi}\left(2 \ln \frac{\sqrt{s}}{m_e}-1\right)$ is the effective bremsstrahlung coupling constant, and $m_e$ is the mass of electron.

The beam spread effect ($S_E$) is considered by
\begin{equation}
  \sigma_{\rm obs}(s)=\int\limits_{\sqrt{s}-5S_E}^{\sqrt{s}+5S_E} d\sqrt{s^\prime}\sigma^{\prime}(s')\frac{1}{\sqrt{2\pi}S_{E}}\exp(\frac{-(\sqrt{s}-\sqrt{s^\prime})^2}{2S^2_E}),
\end{equation}
in which the integral of integration is chosen to include the energy spread effect as whole as possible.

Comparing with the partial width of $J/\psi\to \mu^+\mu^-$, the experimental partial decay width of $J/\psi\to f$ is
\begin{equation}
  \Gamma_f=\mathcal{P}(M^2)(\frac{\mathcal{F}}{M^n})^2 \Gamma_{\mu\mu}\left|1+\mathcal{C} e^{i \phi_{\gamma, 3g}}\right|^2.
\end{equation}
Changing the partial decay width to BF, we have
\begin{equation}
  \mathcal{B}(J/\psi \to f)=\frac{\Gamma_f}{\Gamma_{\mu\mu}} \mathcal{B}(J/\psi\to \mu\mu),
\end{equation}
and
\begin{equation}
  \Gamma_{ee/\mu\mu}=\Gamma \mathcal{B}(J/\psi\to ee/\mu\mu).
\end{equation}

The $\chi^2$ is constructed by incorporating both statistical and systematic uncertainties after considering the correlated and uncorrelated items as~\cite{chifit}
\begin{equation}
  \chi^2=(\Delta\vec{X})^{T}(\mathbf{M}^{-1})(\Delta\vec{X}),
\end{equation}
where $\Delta\vec{X}_{i}=y^\text{obs}_i-y^\text{expect}_i(\vec{\theta})$, $y^\text{obs}$ is the measured cross sections, $y^\text{expect}$ is the expected value calculated from the cross section line shape for each $\sqrt{s}$, and $\vec{\theta}$ represents the fit parameters describing the line shape, that include resonance mass $M$ and width $\Gamma$, the leptonic BFs $\mathcal{B}(J/\psi\to ee)$ and $\mathcal{B}(J/\psi\to \mu\mu)$, the power index of the form factor $n$, relative magnitude $\mathcal{C}$ and relative phase $\phi_{\gamma, 3g}$ between the strong and EM amplitudes, the relative phase between the continuum and EM amplitudes $\phi_{\gamma, \text{cont}}$, the BF of the specific decay process $\mathcal{B}(J/\psi\to f)$, and the beam energy spread $S_E$. Six of them are fixed as shown in Table~\ref{tab:fixedxsfit}. The $\mathbf{M}$ is the covariance matrix. The diagonal elements of $\mathbf{M}$ are the total uncertainties calculated with $\mathbf{M}_{ii}=\mathbf{M}_{{\rm stat},i} + \mathbf{M}_{{\rm sys-corr},i} +\mathbf{M}_{{\rm sys-uncorr},i} + \mathbf{M}^{\sqrt{s}}_{i}$, which include statistical uncertainties, correlated and uncorrelated systematic uncertainties of the cross sections, and uncertainties in energy measurement. Here, $M_\text{sys-corr}$ are the uncertainties obtained by an overall study listed in Sec.~\ref{sec:sys}, tracking, PID, $K_S^0$ reconstruction, and $\mathcal{B}(K_S^0\to\pi^+\pi^-)$, while the other terms are assigned to be $\mathbf{M}_\text{sys-uncorr}$. The off-diagonal elements include correlated systematic uncertainties calculated with $\mathbf{M}_{ij}=\mathbf{M}_{\text{sys-corr},i}\times \mathbf{M}_{\text{sys-corr},j}$. Minimization is performed using the TMinuit package~\cite{James:1975dr}, and errors are estimated using the HESSE algorithm.

\begin{figure*}[htbp]\centering
  \includegraphics[width=0.95\textwidth]{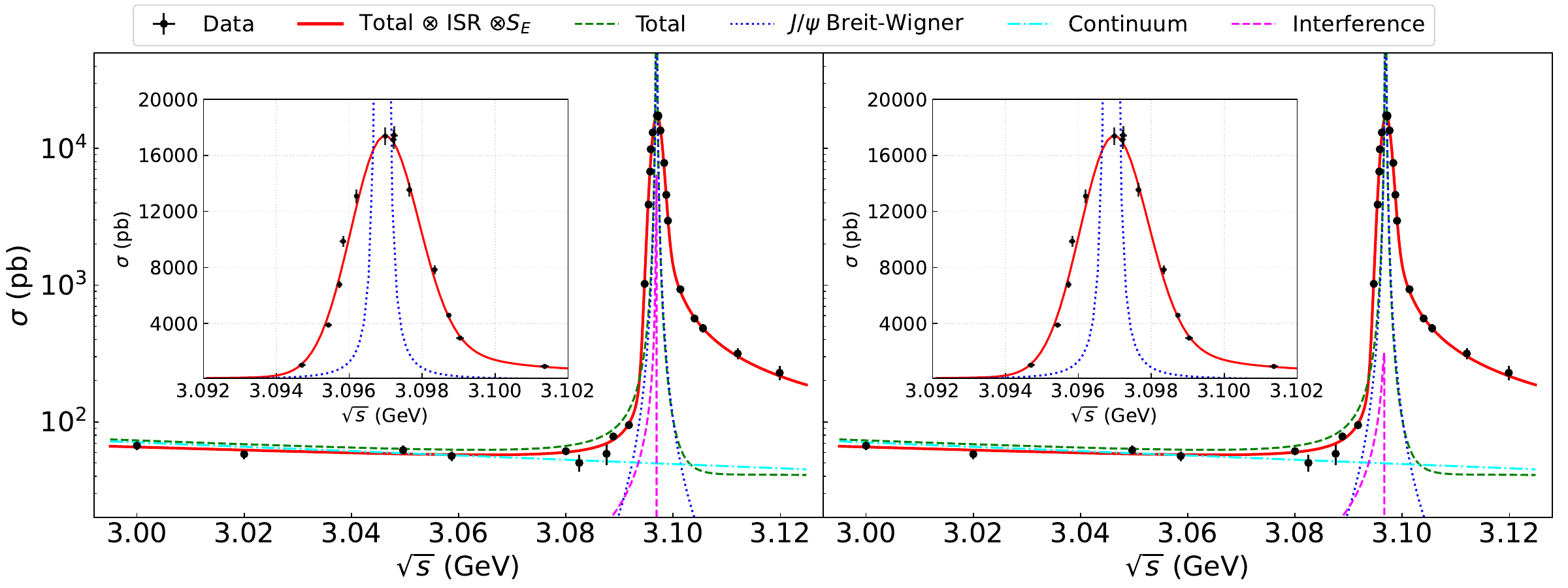}
  \caption{Fits to the observed cross sections of $e^+e^-\to K_S^0 K^+\pi^-$ with positive (left) and negative (right) phase hypotheses.}
  \label{fig:xsfitkskspi}
\end{figure*}

\begin{figure*}[htbp]\centering
  \includegraphics[width=0.95\textwidth]{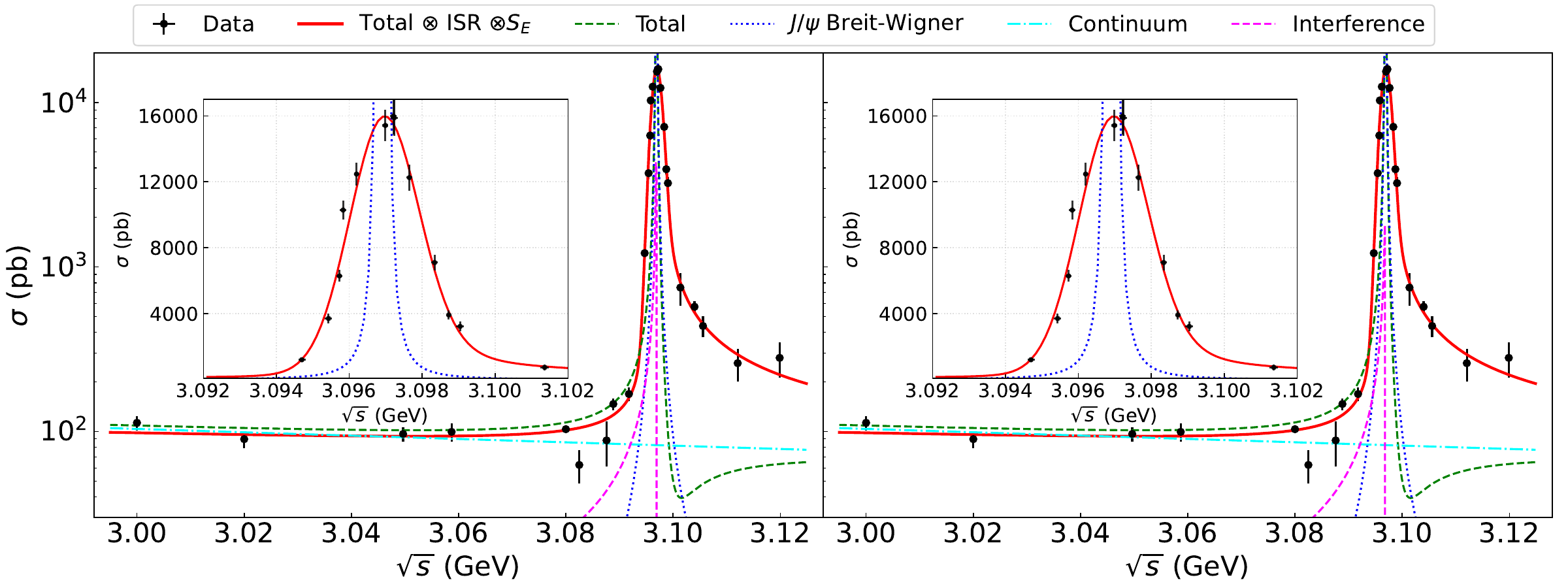}
  \caption{Fits to the observed cross sections of $e^+e^-\to \bar{K}^0 K^*(892)^0$ with positive (left) and negative (right) phase hypotheses.}
  \label{fig:xsfitk0kstar}
\end{figure*}

\begin{figure}[htbp]\centering
  \includegraphics[width=0.48\textwidth]{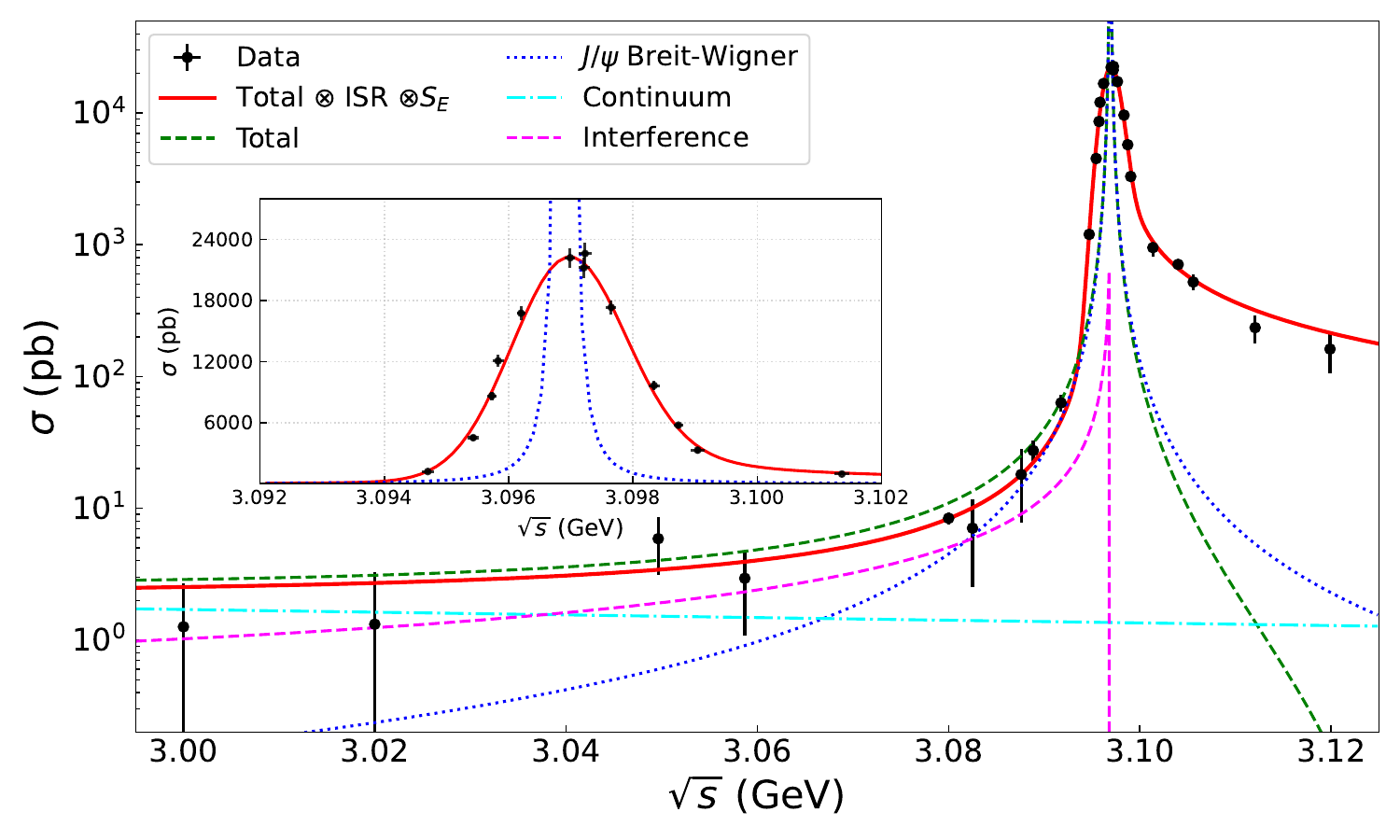}
  \caption{Fits to the observed cross sections of $e^+e^-\to K^+ K^*(892)^-$.}
  \label{fig:xsfitkkstar}
\end{figure}

\begin{figure*}[htbp]\centering
  \includegraphics[width=0.99\textwidth]{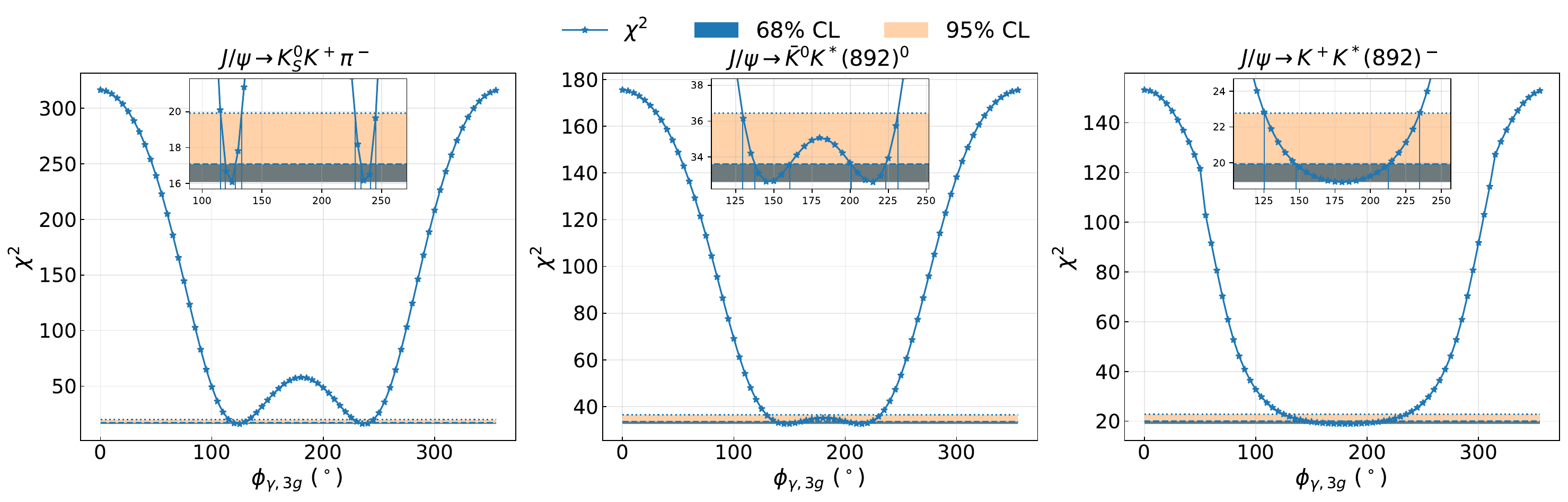}
  \caption{Phase angle scans for the fits to the observed cross sections of (left) $e^+e^-\to K_S^0 K^+\pi^-$, (middle) $e^+e^-\to \bar{K}^0 K^*(892)^0$, and (right) $e^+e^-\to K^+K^*(892)^-$. The dark blue and dark orange bands correspond to the 68\% and 95\% confidence intervals, respectively.}
  \label{fig:scanphase}
\end{figure*}

There are two sets of parameters that describe the cross section equally well. Those with the parameter $\phi_{\gamma,3g}\in(0,180)^\circ$ and $\phi_{\gamma,3g}\in(-180,0)^\circ$ are called positive and negative solution, respectively. Current data cannot distinguish between the positive and negative solutions because the squared amplitude has an intrinsic twofold ambiguity~\cite{Bai:2019jrb}. The best-fit results for $e^+e^-\to K^0_SK^+\pi^-$, $e^+e^-\to \bar{K}^0 K^*(892)^0$, and $e^+e^-\to K^+ K^*(892)^-$ are shown in Figs.~\ref{fig:xsfitkskspi}, \ref{fig:xsfitk0kstar}, and \ref{fig:xsfitkkstar}, respectively, and the fitted parameters are summarized in Table~\ref{tab:xsparakskspi}. The good fit quality in all channels is obtained, and two interference solutions appear for $J/\psi\to K_S^0K^+\pi^-$ and $J/\psi\to \bar K^0K^{*}(892)^0$, while a single solution is found for $J/\psi\to K^{+}K^{*}(892)^-$ (see Table~\ref{tab:xsparakskspi}). The extracted relative magnitude between strong and EM amplitudes in $J/\psi\to K^{+}K^{*}(892)^-$ is very large, indicating a negligible EM contribution and an almost purely strong decay. To determine the confidence interval for the $\phi_{\gamma,3g}$, we perform a scan on this parameter with its value fixed from $0^{\circ}$ to $360^{\circ}$ with a step of $5^{\circ}$, while all other parameters are free and the resulting $\chi^2$ value is shown in Fig.~\ref{fig:scanphase}. The scans are displayed over $(0^\circ,360^\circ)$, whereas the numerical results are quoted in $(-180^\circ,180^\circ)$. The positive phase solutions are unchanged, while the negative phase solutions are displayed after adding one full period, $360^\circ$, make $(-180^\circ,0^\circ)$ mapped onto $(180^\circ,360^\circ)$. The phase scans confirm the single solution for $J/\psi\to K^{+}K^{*}(892)^-$ and disfavor orthogonal or purely real amplitudes at a 3$\sigma$ significance. For the neutral channel, $|\phi_{\gamma,3g}|\simeq(147\pm 12)^\circ$ implies a sizable dispersive strong component opposite in sign to the EM amplitude. The $4.8\sigma$ deviation from $90^\circ$ shows that strong-EM orthogonality is not universal~\cite{ppphase,NNphase,wangyadi,sunzequn}, disfavoring scenarios with a channel-independent, predominantly imaginary strong amplitude, such as a universal vector glueball admixture or absorptive OZI contributions~\cite{gluonball1,gluonball2,OZIamp}, while remaining compatible with FSI dynamics~\cite{FSI1,FSI2,FSIopechar1,FSIopechar2}.

\begin{table*}[htbp]\centering
	\caption{Fixed parameters in the fit to the observed cross sections of $e^+e^-\to K_S^0 K^+\pi^-$ and its subprocesses.}
	\begin{tabular}{c|c|c|c|c|c}
    \toprule
    $M$ (GeV) & $\Gamma$ (keV) & $\mathcal{B}_{J/\psi\to ee}(\%)$ & $\mathcal{B}_{J/\psi\to \mu\mu}(\%)$ & $n$ & $\phi_{\gamma, \text{cont}}$ ($^\circ$) \\
    \midrule
    3.0969 & 92.6 & 5.971 & 5.961 & 5 and 3 & 0 \\
    \bottomrule
    \end{tabular}
	\label{tab:fixedxsfit}
\end{table*}

\begin{table*}[htbp]\centering
  \caption{Best-fit results to the observed cross sections $e^+e^-\to K_S^0 K^+\pi^-$ and its subprocesses. The uncertainties include both statistical and systematic uncertainties.}
  \begin{tabular}{c|c|c|c|c|c|c}
    \toprule
    Channel & $\phi_{\gamma,3g}$ sign & $\mathcal{C}$   & $\mathcal{B}~(\times 10^{-3})$ & $\phi_{\gamma,3g}~(^\circ)$  & $S_E$ (MeV)    & $\chi^2/\text{n.d.f.}$ \\
    \midrule
    \multirow{2}{*}{$J/\psi\to K_S^0K^+\pi^-$} &
    Positive   & $4.31\pm 0.22$  & $5.17\pm 0.20$                 & $ 123.7\pm 5.3$  & $0.90\pm 0.02$ & $16.0/22$           \\
    & Negative & $4.38\pm 0.22$  & $5.36\pm 0.20$                 & $-123.1\pm 5.2$  & $0.90\pm 0.02$ & $16.0/22$           \\
    \midrule
    \multirow{2}{*}{$J/\psi\to\bar{K}^0K^*(892)^0$} &
    Positive   & $3.80\pm 0.26$  & $4.62\pm 0.19$                 & $ 147.0\pm 11.7$ & $0.92\pm 0.03$ & $32.6/22$           \\
    & Negative & $3.85\pm 0.25$  & $4.80\pm 0.19$                 & $-146.0\pm 11.6$ & $0.92\pm 0.03$ & $32.6/22$           \\
    \midrule
    $J/\psi\to K^+K^*(892)^-$ & ...  & $28.66\pm 2.68$ & $6.41 \pm 0.25$     & $180.2\pm 35.2$  & $0.89\pm 0.02$ & $18.9/22$  \\
    \bottomrule
  \end{tabular}
	\label{tab:xsparakskspi}
\end{table*}

The systematic uncertainties on the resonant parameters come from the absolute $\sqrt{s}$ measurement, the measured cross section, and the fit procedure. The uncertainty of the $\sqrt{s}$ measurement and the statistical and systematic uncertainties of the measured cross section are incorporated in the fit; thus, no further uncertainty is necessary. The uncertainties associated with the fit procedure include those arising from the choice of the fit function and the choice of the parameter set. Since the parametrization of the resonance state and continuum is common, the main systematic uncertainties are from the fixed parameters in the fit procedure (see Table~\ref{tab:syslineshapefit}).

\begin{itemize}
  \item \textit{Continuum model}. To estimate the uncertainty due to the continuum model, the alternative fit is performed by increasing or decreasing $n$ by 2.
  \item \textit{Fixed parameters}. Uncertainties associated with those of $M$, $\Gamma$, $\mathcal{B}_{J/\psi\to ee}$, and $\mathcal{B}_{J/\psi\to \mu\mu}$ are evaluated by shifting the corresponding values within $\pm 1\sigma$.
\end{itemize}

\begin{table}[htbp]\centering
  \caption{The systematic uncertainties (in \%) in the fits to the observed cross sections.}
  \begin{tabular}{c|c|c|c|c|c|c}
    \toprule
                  & $M$ & $\Gamma$ & $\mathcal{B}_{J/\psi\to ee}$ & $\mathcal{B}_{J/\psi\to \mu\mu}$ & $n$ & Total \\
    \midrule
    \multicolumn{7}{c}{$J/\psi\to K_S^0K^+\pi^-$}                                                                  \\
    \midrule
    $\mathcal{C}$ & 0.1 & 1.1      & 0.2                          & 0.2                              & 3.8 & 3.9   \\
    $\mathcal{B}$ & 0.0 & 1.9      & 0.5                          & 0.0                              & 0.3 & 2.0   \\
    $\phi$        & 0.2 & 0.3      & 0.0                          & 0.0                              & 2.2 & 2.3   \\
    $S_E$         & 0.0 & 0.1      & 0.0                          & 0.0                              & 0.2 & 0.3   \\
    \midrule
    \multicolumn{7}{c}{$J/\psi\to \bar{K}^0K^*(892)^0$}                                                            \\
    \midrule
    $\mathcal{C}$ & 0.2 & 1.0      & 0.2                          & 0.2                              & 3.4 & 3.6   \\
    $\mathcal{B}$ & 0.0 & 1.9      & 0.5                          & 0.0                              & 0.5 & 2.0   \\
    $\phi$        & 0.3 & 0.4      & 0.0                          & 0.0                              & 3.0 & 3.0   \\
    $S_E$         & 0.0 & 0.1      & 0.0                          & 0.0                              & 0.1 & 0.1   \\
    \midrule
    \multicolumn{7}{c}{$J/\psi\to K^+K^*(892)^-$}                                                                  \\
    \midrule
    $\mathcal{C}$ & 0.1 & 0.5      & 0.3                          & 0.3                              & 2.3 & 2.4   \\
    $\mathcal{B}$ & 0.1 & 1.9      & 0.5                          & 0.0                              & 0.0 & 1.9   \\
    $\phi$        & 0.0 & 0.0      & 0.0                          & 0.0                              & 0.0 & 0.1   \\
    $S_E$         & 0.0 & 0.1      & 0.0                          & 0.0                              & 0.0 & 0.1   \\
    \bottomrule
  \end{tabular}
  \label{tab:syslineshapefit}
\end{table}

\section{ISOSPIN AND DIRECT $CP$ ASYMMETRY}

To obtain the direct $CP$ asymmetries, we perform an additional simultaneous $\chi^2$ fit to ``$+$" and ``$-$" charge conjugate data samples. The residual vector $\Delta\vec{X}$ is defined as the concatenation of the two charge samples,
\begin{equation}
  \Delta\vec{X} =
  \begin{pmatrix}
    \Delta\vec{X}^{+} \\
    \Delta\vec{X}^{-}
  \end{pmatrix}
  =
  \begin{pmatrix}
    y^{\text{obs},+}_i - y^{\text{exp},+}_i(\vec{\theta}^+) \\
    y^{\text{obs},-}_j - y^{\text{exp},-}_j(\vec{\theta}^-)
  \end{pmatrix},
\end{equation}
where $y^{\text{obs},\pm}_i$ and $y^{\text{exp},\pm}_i(\vec{\theta}^\pm)$ are, respectively, the measured observed cross sections and the corresponding expected cross sections from the line shape parametrization at each c.m. energy point. The parameter vectors $\vec{\theta}^{+}$ and $\vec{\theta}^{-}$ contain three charge-dependent parameters, the relative magnitude $\mathcal{C}^{\pm}$, the phase $\phi_{\gamma,3g}^{\pm}$, and the BF $\mathcal{B}^{\pm}\equiv \mathcal{B}(J/\psi\to f^\pm)$, while all other quantities describing the resonance and continuum properties are common to the two charged samples. In particular, the resonance mass $M$, total width $\Gamma$, leptonic BFs $\mathcal{B}(J/\psi\to ee)$ and $\mathcal{B}(J/\psi\to \mu\mu)$, the power index $n$ of the continuum amplitude, and the phase $\phi_{\gamma,\text{cont}}$ are shared parameters in the simultaneous fit.

The covariance matrix is constructed in a similar way, with the full covariance matrix $\mathbf{M}$ defined as a $52\times52$ block matrix:
\begin{equation}
  \mathbf{M} =
  \begin{pmatrix}
    \mathbf{M}^{++} & \mathbf{M}^{+-} \\
    \mathbf{M}^{-+} & \mathbf{M}^{--}
  \end{pmatrix},
\end{equation}
where the diagonal blocks $\mathbf{M}^{++}$ and $\mathbf{M}^{--}$ are the covariance matrices of the $+$ and $-$ samples, respectively. They are constructed in the same way as for the nominal line shape fit, with the diagonal elements containing the quadratic sum of the statistical uncertainties, the uncorrelated systematic uncertainties, and the contribution from the c.m. energy uncertainty $\mathbf{M}^{\pm\pm}_{ii}=\mathbf{M}^{\pm}_{{\rm stat},i} + \mathbf{M}^{\pm}_{{\rm sys-corr},i} +\mathbf{M}^{\pm}_{{\rm sys-uncorr},i} + \mathbf{M}^{\pm,\sqrt{s}}_{i}$, while the off-diagonal elements in each block encode the correlations caused by the common systematic sources $\mathbf{M}^{\pm\pm}_{ij}=\mathbf{M}^{\pm}_{\text{sys-corr},i}\times \mathbf{M}^{\pm}_{\text{sys-corr},j}$. Here, $\mathbf{M}^{\pm}_{\text{sys-corr}}$ includes the correlated systematic uncertainties.

The off-diagonal blocks $\mathbf{M}^{+-}$ and $\mathbf{M}^{-+}$ represent the correlations between the $+$ and $-$ charge samples, where the systematic uncertainties of the cross sections and uncertainties in energy measurement are considered fully correlated between the two charged samples. From the simultaneous fit we form the charge asymmetries of the line shape parameters, $A_{\mathcal{C}}$, $A_{\mathcal{B}}$, and $\Delta\phi_{\gamma,3g}$. The uncertainties on $A_p$ ($p\in\{\mathcal{C},\mathcal{B}\}$) are evaluated by propagating the full covariance matrix of the fitted parameters. In particular, for each pair $(p^{+},p^{-})$ we use the $2\times 2$ submatrix of the covariance matrix:
\begin{equation}\footnotesize
  \sigma^2(A_p) =
  \left(\frac{\partial A_p}{\partial p^{+}},\,
        \frac{\partial A_p}{\partial p^{-}}\right)
  \begin{pmatrix}
    \mathrm{Var}(p^{+}) & \mathrm{Cov}(p^{+},p^{-})\\[2pt]
    \mathrm{Cov}(p^{+},p^{-}) & \mathrm{Var}(p^{-})
  \end{pmatrix}
  \begin{pmatrix}
    \frac{\partial A_p}{\partial p^{+}}\\[2pt]
    \frac{\partial A_p}{\partial p^{-}}
  \end{pmatrix}
\end{equation}
and $\sigma^2(\Delta\phi_{\gamma,3g})=\mathrm{Var}(\phi^+_{\gamma,3g})+\mathrm{Var}(\phi^-_{\gamma,3g})-2\mathrm{Cov}(\phi^+_{\gamma,3g},\phi^-_{\gamma,3g})$, which fully accounts for the strong correlations between the $+$ and $-$ parameters induced by the common systematic in the simultaneous fit. The uncertainties on energy-dependent direct $CP$ asymmetries $A_{\sigma_{\rm Born}}$ are again obtained by propagating the full parameter covariance matrix to these derived observables. Because most of the dominant systematic effects are common to the two charge samples, they largely cancel in the asymmetry observables, and the resulting uncertainties are dominated by the statistical precision and the small residual uncorrelated systematic contributions. It is also worth noting that, in asymmetry parameters, the systematic uncertainty of the fixed parameters in the cross section fitting is negligible.

\begin{figure}[htbp]\centering
  \includegraphics[width=0.48\textwidth]{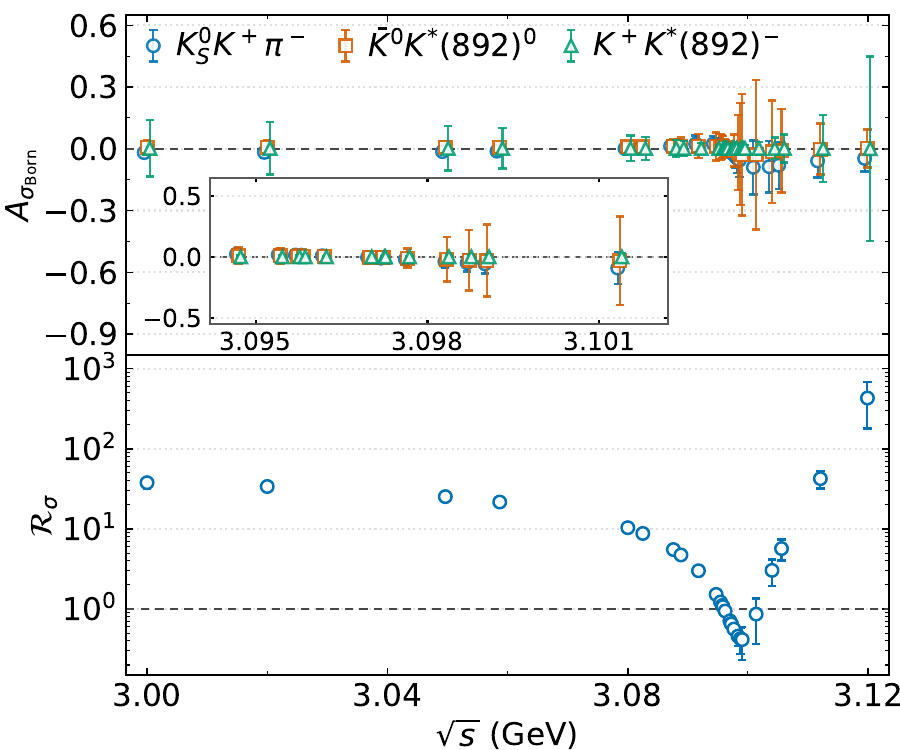}
  \caption{Asymmetry observables of the Born cross sections (top), and the isospin ratio $\mathcal{R}_\sigma$ of the Born cross sections (bottom).}
  \label{fig:asy}
\end{figure}

Similarly, the isospin ratio is extracted with a simultaneous $\chi^2$ fit described in $CP$ asymmetry. The only difference is that, at the same c.m. energy, the statistical uncertainties associated with the event yields are treated as correlated. The isospin-breaking effect is directly reflected in the measured ratio $\mathcal{R}_{K^*\bar{K}}=0.723\pm 0.013$ or $0.752\pm 0.014$. When only the strong amplitude is considered, the ratio increases to $\mathcal{R}^{3g}_{K^*\bar{K}}=1.062\pm 0.070$ or $1.090\pm 0.063$, showing the isospin symmetry is conserved in strong interaction. In contrast, the contribution from the EM amplitude alone yields a much larger ratio of $\mathcal{R}^{\gamma}_{K^*\bar{K}}=62.0\pm 12.2$, indicating a very pronounced isospin-breaking effect from the EM interaction, while far from the $J/\psi$ resonance the ratio approaches $\sim37.74\pm 5.57$, again showing sizable isospin breaking in the energy regions dominated by the continuum. The $\mathcal{C}=3.80\pm 0.26$ versus $28.66\pm 2.68$ corresponds to $|\mathcal{A}_{\gamma}/\mathcal{A}_{3g}|=0.263\pm 0.018$ and $0.035\pm 0.003$ in the neutral and charged modes, respectively. Together with $\mathcal{R}^{3g}_{K^*\bar K}$ and $\mathcal{R}^{\gamma}_{K^*\bar K}$, these results indicate that the isospin breaking resides mainly in the EM amplitude, while the strong contribution remains isospin symmetric. This qualitatively supports effective Lagrangian expectations of strong dominance and similar strong flavor-breaking effects in both modes~\cite{VPEFT}, while the pronounced observed EM amplitude constrains the channel-dependent EM operators and possible hadronic loop effects~\cite{FSIopechar1,FSIopechar2}. The results are listed in Table~\ref{tab:iso_rat} and shown in the bottom figure in Fig.~\ref{fig:asy}. The resulting asymmetries $A_{\sigma_{\rm Born}}$ are shown in the top figure in Fig.~\ref{fig:asy} and listed in Table~\ref{tab:energy_asym}, while the parameter asymmetries $A_{\mathcal{C}}$, $A_{\mathcal{B}}$, and $\Delta\phi_{\gamma,3g}$ are summarized in Table~\ref{tab:param_asym}. All measured null-test observables are consistent with the Standard Model expectation of zero within one standard deviation. The $CP$ results are quoted for the positive phase solution; the negative phase solution gives the same $CP$-null conclusion.

\begin{table}[htbp]\centering
  \caption{The isospin ratio $\mathcal{R}_\sigma$ of the Born cross section.}
  \begin{tabular}{cccc}
    \toprule
    $\sqrt{s}$ (\si{MeV}) & $\mathcal{R}_\sigma$ \\
    \midrule
    $3000.00 \pm 0.20$    & $37.74 \pm 5.57$ \\
    $3020.00 \pm 0.20$    & $33.84 \pm 4.69$ \\
    $3049.64 \pm 0.06$    & $25.29 \pm 2.95$ \\
    $3058.69 \pm 0.06$    & $21.66 \pm 2.30$ \\
    $3080.00 \pm 0.20$    & $10.35 \pm 0.71$ \\
    $3082.50 \pm 0.06$    & $8.77 \pm 0.56$ \\
    $3087.59 \pm 0.13$    & $5.52 \pm 0.33$ \\
    $3088.85 \pm 0.06$    & $4.73 \pm 0.29$ \\
    $3091.76 \pm 0.06$    & $3.00 \pm 0.22$ \\
    $3094.70 \pm 0.10$    & $1.52 \pm 0.12$ \\
    $3095.43 \pm 0.10$    & $1.22 \pm 0.09$ \\
    $3095.73 \pm 0.08$    & $1.11 \pm 0.08$ \\
    $3095.83 \pm 0.09$    & $1.07 \pm 0.07$ \\
    $3096.20 \pm 0.07$    & $0.94 \pm 0.05$ \\
    $3096.99 \pm 0.08$    & $0.71 \pm 0.02$ \\
    $3097.21 \pm 0.09$    & $0.66 \pm 0.03$ \\
    $3097.23 \pm 0.10$    & $0.65 \pm 0.03$ \\
    $3097.65 \pm 0.08$    & $0.56 \pm 0.06$ \\
    $3098.34 \pm 0.09$    & $0.46 \pm 0.11$ \\
    $3098.73 \pm 0.08$    & $0.43 \pm 0.15$ \\
    $3099.04 \pm 0.11$    & $0.42 \pm 0.18$ \\
    $3101.36 \pm 0.12$    & $0.86 \pm 0.50$ \\
    $3104.00 \pm 0.08$    & $3.04 \pm 1.10$ \\
    $3105.58 \pm 0.10$    & $5.68 \pm 1.67$ \\
    $3112.05 \pm 0.11$    & $42.44 \pm 10.63$ \\
    $3119.88 \pm 0.13$    & $430.43 \pm 249.94$ \\
    \bottomrule
  \end{tabular}
  \label{tab:iso_rat}
\end{table}

\begin{table*}[htbp]\centering
  \caption{The asymmetry observable $A_{\sigma_\text{Born}}$ for $e^+e^-\to K_S^0 K^+\pi^-$ and its subprocesses.}
  \begin{tabular}{cccc}
    \toprule
    $\sqrt{s}$ (\si{MeV}) & $e^+e^-\to K_S^0K^+\pi^-$        & $e^+e^-\to \bar K^0 K^*(892)^0$  & $e^+e^-\to K^+ K^*(892)^-$      \\
    \midrule
    $3000.00 \pm 0.20$ & $-0.015\,\pm\,0.024$ & $ 0.007\,\pm\,0.035$ & $0.002\,\pm\,0.137$ \\
    $3020.00 \pm 0.20$ & $-0.014\,\pm\,0.023$ & $ 0.007\,\pm\,0.033$ & $0.002\,\pm\,0.128$ \\
    $3049.64 \pm 0.06$ & $-0.011\,\pm\,0.019$ & $ 0.008\,\pm\,0.027$ & $0.002\,\pm\,0.108$ \\
    $3058.69 \pm 0.06$ & $-0.009\,\pm\,0.017$ & $ 0.008\,\pm\,0.025$ & $0.002\,\pm\,0.099$ \\
    $3080.00 \pm 0.20$ & $ 0.002\,\pm\,0.015$ & $ 0.010\,\pm\,0.020$ & $0.002\,\pm\,0.062$ \\
    $3082.50 \pm 0.06$ & $ 0.005\,\pm\,0.017$ & $ 0.010\,\pm\,0.023$ & $0.002\,\pm\,0.056$ \\
    $3087.59 \pm 0.13$ & $ 0.014\,\pm\,0.026$ & $ 0.012\,\pm\,0.037$ & $0.002\,\pm\,0.041$ \\
    $3088.85 \pm 0.06$ & $ 0.016\,\pm\,0.030$ & $ 0.012\,\pm\,0.042$ & $0.002\,\pm\,0.036$ \\
    $3091.76 \pm 0.06$ & $ 0.023\,\pm\,0.038$ & $ 0.013\,\pm\,0.058$ & $0.002\,\pm\,0.025$ \\
    $3094.70 \pm 0.10$ & $ 0.023\,\pm\,0.037$ & $ 0.012\,\pm\,0.068$ & $0.002\,\pm\,0.013$ \\
    $3095.43 \pm 0.10$ & $ 0.019\,\pm\,0.030$ & $ 0.010\,\pm\,0.061$ & $0.002\,\pm\,0.010$ \\
    $3095.73 \pm 0.08$ & $ 0.016\,\pm\,0.026$ & $ 0.009\,\pm\,0.055$ & $0.002\,\pm\,0.009$ \\
    $3095.83 \pm 0.09$ & $ 0.015\,\pm\,0.024$ & $ 0.009\,\pm\,0.053$ & $0.002\,\pm\,0.009$ \\
    $3096.20 \pm 0.07$ & $ 0.011\,\pm\,0.018$ & $ 0.007\,\pm\,0.041$ & $0.002\,\pm\,0.008$ \\
    $3096.99 \pm 0.08$ & $-0.002\,\pm\,0.004$ & $ 0.001\,\pm\,0.010$ & $0.002\,\pm\,0.007$ \\
    $3097.21 \pm 0.09$ & $-0.006\,\pm\,0.010$ & $-0.002\,\pm\,0.028$ & $0.002\,\pm\,0.008$ \\
    $3097.23 \pm 0.10$ & $-0.006\,\pm\,0.011$ & $-0.002\,\pm\,0.030$ & $0.002\,\pm\,0.008$ \\
    $3097.65 \pm 0.08$ & $-0.016\,\pm\,0.025$ & $-0.008\,\pm\,0.078$ & $0.002\,\pm\,0.009$ \\
    $3098.34 \pm 0.09$ & $-0.033\,\pm\,0.051$ & $-0.019\,\pm\,0.183$ & $0.002\,\pm\,0.012$ \\
    $3098.73 \pm 0.08$ & $-0.042\,\pm\,0.065$ & $-0.025\,\pm\,0.248$ & $0.002\,\pm\,0.014$ \\
    $3099.04 \pm 0.11$ & $-0.049\,\pm\,0.076$ & $-0.029\,\pm\,0.296$ & $0.002\,\pm\,0.016$ \\
    $3101.36 \pm 0.12$ & $-0.080\,\pm\,0.123$ & $-0.028\,\pm\,0.363$ & $0.002\,\pm\,0.031$ \\
    $3104.00 \pm 0.08$ & $-0.077\,\pm\,0.117$ & $-0.014\,\pm\,0.248$ & $0.002\,\pm\,0.053$ \\
    $3105.58 \pm 0.10$ & $-0.071\,\pm\,0.107$ & $-0.010\,\pm\,0.204$ & $0.002\,\pm\,0.068$ \\
    $3112.05 \pm 0.11$ & $-0.052\,\pm\,0.077$ & $-0.001\,\pm\,0.124$ & $0.002\,\pm\,0.162$ \\
    $3119.88 \pm 0.13$ & $-0.041\,\pm\,0.060$ & $ 0.002\,\pm\,0.092$ & $0.001\,\pm\,0.448$ \\
    \bottomrule
  \end{tabular}
  \label{tab:energy_asym}
\end{table*}

\begin{table}[htbp]\centering
  \caption{Parameter asymmetries for $\mathcal{C}$, $\mathcal{B}$ and $\phi_{\gamma,3g}$ in cross section fits.}
  \resizebox{\linewidth}{!}{
    \begin{tabular}{cccc}
      \toprule
      Channel                         & $A_{\mathcal C}$ & $A_{\mathcal B}$& $\Delta\phi_{\gamma,3g}~(^\circ)$  \\
      \midrule
      $J/\psi\to K_S^0K^+\pi^-$       & $0.018\pm 0.027$  & $0.001\pm 0.003$  & $4.7\pm 7.3$   \\
      $J/\psi\to \bar{K}^0K^*(892)^0$ & $0.001\pm 0.048$  & $0.002\pm 0.009$  & $1.7\pm 15.8$  \\
      $J/\psi\to K^+K^*(892)^-$       & $-0.000\pm 0.087$ & $0.002\pm 0.008$  & $-0.0\pm 66.7$ \\
      \bottomrule
    \end{tabular}}
  \label{tab:param_asym}
\end{table}

\section{SUMMARY}
Using $e^+e^-$ annihilation data collected at 26 c.m. energy points between 3000.00 and $3119.88~\text{MeV}$ with the BESIII detector, corresponding to a total integrated luminosity of about $440.7~\text{pb}^{-1}$, we study the line shape of $e^+e^-\to K_S^0 K^+\pi^-$. Based on the PWA results, we further investigate the line shapes of the dominant subprocesses $e^+e^-\to \bar K^0 K^*(892)^0$ and $e^+e^-\to K^+ K^*(892)^-$. Coherent line shape fits determine the relative strong-EM phase $\phi_{\gamma,3g}$, magnitude $\mathcal C$, and the corresponding BF $\mathcal B$. For $(\phi_{\gamma,3g},\mathcal C,\mathcal B)$, the two solutions for $J/\psi\to K_S^0K^+\pi^-$ are $((123.7\pm 5.3)^\circ,\,4.31\pm 0.22,\,(5.17\pm 0.20)\times 10^{-3})$ or $((-123.1\pm 5.2)^\circ,\,4.38\pm 0.22,\,(5.36\pm 0.20)\times 10^{-3})$. For $J/\psi\to\bar K^0K^*(892)^0$, the two solutions are $((147.0\pm 11.7)^\circ,\,3.80\pm 0.26,\,(4.62\pm 0.19)\times 10^{-3})$ or $((-146.0\pm 11.6)^\circ,\,3.85\pm 0.25,\,(4.80\pm 0.19)\times 10^{-3})$, while $J/\psi\to K^+K^*(892)^-$ has the single solution $((180.2\pm 35.2)^\circ,\,28.66\pm 2.68,\,(6.41 \pm 0.25)\times 10^{-3})$. All uncertainties include statistical and systematic contributions. The BFs are consistent with the world averages and have substantially improved precision.

For the decay of $J/\psi\to\bar K^0 K^*(892)^0$, the extracted phase deviates from the orthogonal relation by $4.8\sigma$ and differs from the pQCD prediction of $0^\circ$ by $12.6\sigma$ and from $180^\circ$ by $2.8\sigma$. For the decay of $J/\psi\to K^+ K^*(892)^-$, the central value is located at $180^\circ$, but it still deviates from the orthogonal relation by $2.6\sigma$ and from $0^\circ$ by $5.1\sigma$. Given these unexpected values and the apparent channel dependence, the phase behavior appears more compatible with channel-dependent FSI or hadronic loop triangle effects than with a channel-independent glueball mixing scenario.

The isospin ratios are measured to be $\mathcal{R}_{K^*\bar{K}}=0.723\pm 0.013$ or $0.752\pm 0.014$. The ratios obtained from the strong amplitudes alone are $\mathcal{R}^{3g}_{K^*\bar{K}}=1.062\pm 0.070$ or $1.090\pm 0.063$, which deviate from unity by $0.9\sigma$ and $1.4\sigma$, respectively. This is broadly consistent with isospin conservation in the strong interaction. In contrast, the EM contribution alone yields a much larger ratio of $\mathcal{R}^{\gamma}_{K^*\bar{K}}=62.0\pm 12.2$, indicating a very pronounced isospin-breaking effect from the EM interaction. Far from the $J/\psi$ resonance, the ratio approaches $\sim38$ at $\sqrt{s}=3000.00~\text{MeV}$, with a $2.0\sigma$ deviation from the $\mathcal{R}^{\gamma}_{K^*\bar{K}}$.

Finally, the direct $CP$ null tests are performed, and no evidence for $CP$ violation is observed, either at the Born cross section level or in the amplitude parameters $\mathcal{C}$, $\phi_{\gamma,3g}$, and $\mathcal{B}$. All results are consistent with zero within 1$\sigma$.

\textbf{Acknowledgement}

The BESIII Collaboration thanks the staff of BEPCII (https://cstr.cn/31109.02.BEPC) and the IHEP computing center for their strong support. This work is supported in part by National Key R\&D Program of China under Contracts Nos. 2023YFA1606000, 2023YFA1606704, 2025YFA1613900; National Natural Science Foundation of China (NSFC) under Contracts Nos. 11635010, 11935015, 11935016, 11935018, 12025502, 12035009, 12035013, 12061131003, 12192260, 12192261, 12192262, 12192263, 12192264, 12192265, 12221005, 12225509, 12235017, 12342502, 12361141819, 12535005; the Chinese Academy of Sciences (CAS) Large-Scale Scientific Facility Program; the Strategic Priority Research Program of Chinese Academy of Sciences under Contract No. XDA0480600; CAS under Contract No. YSBR-101; 100 Talents Program of CAS; The Institute of Nuclear and Particle Physics (INPAC) and Shanghai Key Laboratory for Particle Physics and Cosmology; ERC under Contract No. 758462; German Research Foundation DFG under Contract No. FOR5327; Istituto Nazionale di Fisica Nucleare, Italy; Knut and Alice Wallenberg Foundation under Contracts Nos. 2021.0174, 2021.0299, 2023.0315; Ministry of Development of Turkey under Contract No. DPT2006K-120470; National Research Foundation of Korea under Contract No. NRF-2022R1A2C1092335; National Science and Technology fund of Mongolia; Polish National Science Centre under Contract No. 2024/53/B/ST2/00975; STFC (United Kingdom); Swedish Research Council under Contract No. 2019.04595; U. S. Department of Energy under Contract No. DE-FG02-05ER41374

The main results of this study are also reported in the companion Letter~\cite{PRLcompanion}.

\textbf{DATA AVAILABILITY}

The data that support the findings of this article are not publicly available upon publication because it is not technically feasible and/or the cost of preparing, depositing, and hosting the data would be prohibitive within the terms of this research project. The data are available from the authors upon reasonable request.

\end{document}